\documentclass[journal,twocolumn]{IEEEtran-v17}

\newlength{\figwidth}

\usepackage[nosort]{cite}
\usepackage{url}
\usepackage[intlimits]{amsmath}
\usepackage{bbm}
 \usepackage[dvips]{graphicx}
   \DeclareGraphicsExtensions{.eps,.pdf}

\usepackage{paralist}
\usepackage{fancyref}
\usepackage[stretch=16,shrink=16,step=4]{microtype}
\usepackage{pdfsync}
\usepackage{vmr-symbols-vecbold}
\usepackage{standard-macros}
\usepackage{ifthen}
\usepackage{colortbl} 

\newboolean{arxiv}
\setboolean{arxiv}{true} 

\ifthenelse {\boolean{arxiv}}
{
}
{
\setlength{\figwidth}{0.65\textwidth}
\usepackage{color}
\definecolor{links}{rgb}{0.7,0,0}   
\definecolor{urls}{rgb}{0,0,0.8}    
\definecolor{cites}{rgb}{0,0,0.8}   

\usepackage[colorlinks,hyperindex,linkcolor=links,citecolor=cites,urlcolor=urls]{hyperref} 
}

\makeatletter
\def\@IEEEinterspaceratioM{0.265}
\def\@IEEEinterspaceMINratioM{0.1651}
\def\@IEEEinterspaceMAXratioM{0.38}

\def\@IEEEinterspaceratioB{0.31}
\def\@IEEEinterspaceMINratioB{0.19}
\def\@IEEEinterspaceMAXratioB{0.38}
\@IEEEtunefonts
\makeatother
\safemath{\cohtime}{T}               	
\safemath{\rxant}{N}                    
\safemath{\txant}{M}                    
\safemath{\inpdist}{\mathsf{Q}_\matX}   
\safemath{\txantopt}{\txant^*}			

\safemath{\minparam}{\underline{L}}					
\safemath{\maxparam}{\overline{L}}                 
\safemath{\minbound}{\underline{P}}					
\safemath{\maxbound}{\overline{P}}                    

\safemath{\varsnr}{\lambda}				

\safemath{\const}{c}	
\safemath{\constrho}{c_{\rho_0}}	
\safemath{\constcm}{c\sub{USTM}}										
\safemath{\grassman}{\setG}							
\safemath{\stiefel}{\setS}							
\safemath{\substiefel}{\widetilde{\stiefel}}		

\safemath{\chosoutdist}{\mathsf{R}_{\matY}}   			
\safemath{\condoutdist}{\mathsf{P}_{\matY\given\matX}}   
\safemath{\condoutpdf}{\mathsf{p}}                       
\safemath{\chosoutpdf}{\mathsf{r}_\matY}                    
\safemath{\outpdf}{\mathsf{r}}
\safemath{\outdist}{\mathsf{Q}_{\matY}}				  
\safemath{\genericpdf}{\mathsf{f}}					   
\safemath{\altgenericpdf}{\mathsf{p}}				   

\newcommand{\biprod}[2]{\ensuremath{\prod\limits_{#1}^{#2}}}  

\newcommand{\eig}[2]{\lambda_{#1}\lefto\{#2 \right\}}								

\safemath{\auxdiagmat}{\widetilde{\matD}}      
\safemath{\auxdiagent}{\tilde{d}}              

\safemath{\svdleftalt}{\widetilde{\matU}}		
\safemath{\svdrightalt}{\widetilde{\matV}}      
\safemath{\svddiagalt}{\widetilde{\matSigma}}   

\safemath{\rotmat}{\matP}						
\safemath{\altrotmat}{\widetilde{\matP}}					

\safemath{\outmatsmall}{\widetilde{\matY}}		

\DeclareMathOperator{\betadist}{Beta}			

\safemath{\custm}{C\sub{USTM}}          		

\safemath{\approxc}{\widetilde{C}}              
\safemath{\approxcustm}{\widetilde{C}\sub{USTM}}   

\safemath{\wishart}{\mathcal{W}}					

\safemath{\distoptD}{\mathsf{Q}^{\text{opt}}_{\matD}} 

\safemath{\distoptDpar}{\mathsf{Q}^{\text{opt},\snr}_{\matD}} 

\safemath{\pdfoptD}{\mathsf{q}^{\text{opt}}_{\matD}}  

\safemath{\pdfoptDpar}{\mathsf{q}^{\text{opt},\snr}_{\matD}}  

\safemath{\altsigma}{\tilde{\sigma}}      				

\safemath{\vecsigma}{\bm\sigma}              			
\safemath{\altvecsigma}{\tilde{\vecsigma}}              
\safemath{\altvecsigmaarg}{\veca}
\safemath{\matDarg}{\matLambda}

\safemath{\altnoisemat}{\widetilde{\matW}}              

\safemath{\prelog}{\chi}                                

\safemath{\pdfaltsigma}{\genericpdf_{\altvecsigma}^{(\snr)}}       
\safemath{\pdfu}{\genericpdf_{\vecu}}                                         
\safemath{\pdfaltsigmacond}{\genericpdf_{\altvecsigma\given \matD}^{(\snr)}}    
\safemath{\pdfucond}{\genericpdf_{\vecu\given \matD}}                        

\safemath{\pdfaltsigmascalar}{\genericpdf_{\altsigma_i}^{(\snr)}}

\safemath{\altsnr}{\tilde{\snr}}						

\safemath{\altmatSigma}{\matDelta}

\safemath{\unitgroup}{\setU}

\safemath{\altmatPhi}{\widetilde{\matPhi}} 		

\safemath{\altdelta}{\tilde{\delta}}
\safemath{\altlambda}{\tilde{\lambda}}

\safemath{\constapp}{k}			

\safemath{\snrnorm}{\altsnr}
\safemath{\snrth}{\snr\sub{th}}

\safemath{\pdfuscalar}{\genericpdf_{u_i}}
\safemath{\diagsetK}{\widetilde{\setK}}

\begin{document}

\IEEEoverridecommandlockouts
%

\title{On the Capacity of Large-MIMO Block-Fading Channels}
%
%
\author{Wei~Yang,~\IEEEmembership{Student Member,~IEEE}, Giuseppe~Durisi,~\IEEEmembership{Senior Member,~IEEE}, Erwin~Riegler,~\IEEEmembership{Member,~IEEE}

\thanks{
The work of Erwin Riegler was supported by the WWTF under grant ICT10-066 (NOWIRE).}
\thanks{The material in this paper was presented in part at the IEEE International Symposium on Information Theory (ISIT), Boston, MA, July~2012.}

\thanks{W. Yang and G. Durisi are with the Department of Signals and Systems, Chalmers University of Technology, Gothenburg, Sweden (e-mail: \{ywei, durisi\}@chalmers.se).}
\thanks{E. Riegler is with the Institute of Telecommunications, Vienna University of Technology, Vienna, Austria (e-mail: erwin.riegler@nt.tuwien.ac.at).}
}%
\maketitle

\begin{abstract}
 	
  We characterize the capacity of Rayleigh block-fading multiple-input multiple-output (MIMO) channels in the noncoherent setting where transmitter and receiver have no \emph{a priori} knowledge of the realizations of the fading channel.
We prove that \emph{unitary space-time modulation} (USTM) is not capacity-achieving in the high signal-to-noise ratio (SNR) regime when the total number of antennas exceeds the coherence time of the fading channel (expressed in multiples of the symbol duration), a situation that is relevant for MIMO systems with large antenna arrays (large-MIMO systems).
This result settles a conjecture by Zheng \& Tse (2002) in the affirmative.
The capacity-achieving input signal, which we refer to as \emph{Beta-variate space-time modulation} (BSTM), turns out to be the product of a unitary isotropically distributed random matrix, and a diagonal matrix whose nonzero entries are distributed as the square-root of the eigenvalues of a Beta-distributed random matrix of appropriate size.
Numerical results illustrate that using BSTM instead of USTM in large-MIMO systems yields a rate gain as large as $13\%$ for SNR values of practical interest.
\end{abstract}

\section{Introduction} 
\label{sec:introduction}
The use of multiple antennas increases tremendously the throughput of wireless systems operating over fading channels~\cite{foschini98-a,telatar99-11a}.
Specifically, when a genie provides the receiver with perfect channel state information
 (the so called \emph{coherent setting}), the capacity of a multiple-input multiple-output (MIMO) fading channel grows linearly in the minimum between the number of transmit and receive antennas~\cite{telatar99-11a}.
In practice, however, the fading channel
 is not known \emph{a priori} at the receiver and must be estimated, for example through the transmission of pilot symbols.
Lack of \emph{a priori} channel knowledge
 at the receiver determines a capacity loss compared to the coherent case.
This loss, which  depends on the rate at which the fading channel varies in time, frequency, and space~\cite{marzetta99-01a,zheng02-02a,lapidoth06-02a,schuster09-09a}, can be characterized in a
 fundamental way by studying capacity in the \emph{noncoherent setting} where neither the transmitter nor the receiver are assumed to have \emph{a priori} knowledge of the realizations of the fading channel (but both are assumed to know its statistics perfectly).
In the remainder of the paper, we will refer to capacity in the noncoherent setting simply as capacity.
We emphasize that in the noncoherent setting the receiver is allowed to try and gain channel knowledge.
Channel estimation is simply viewed as a specific form of coding~\cite{lapidoth05-07a}.

For frequency-flat fading channels, a simple model to capture channel variations in time is the Rayleigh block-fading model according to which the channel  remains constant over a block of $\cohtime > 1$ symbols and changes independently from block to block.
The parameter \cohtime can be thought of as the channel's coherence time.
Even if the capacity of the Rayleigh block-fading MIMO channel has been studied extensively in the literature~\cite{marzetta99-01a,hochwald00-03a,zheng02-02a,hassibi02-06a}, no closed-form capacity expression is available to date.
Zheng and Tse~\cite{zheng02-02a} showed that capacity behaves in the high signal-to-noise ratio (SNR) regime as\footnote{When $\cohtime =1$, capacity grows double-logarithmically in \snr~\cite[Thm.~4.2]{lapidoth03-10a}.}
\begin{align}\label{eq:zheng_tse_v1}
	C(\snr)=\txantopt\left(1-\frac{\txantopt}{\cohtime}\right)\log( \snr) + \landauO(1), \quad \snr \to \infty.
\end{align}
Here, \snr denotes the SNR, $\txantopt\define\min\{\txant,\rxant, \floor{\cohtime/2}\}$ with \txant and \rxant standing for the number of transmit and receive antennas, respectively, and $\landauO(1)$ indicates a bounded function of \snr (for sufficiently large \snr).
The high-SNR capacity expression given in~\eqref{eq:zheng_tse_v1} is insightful as it allows one to determine the capacity loss (at high SNR) due to lack of \emph{a priori} channel knowledge.
Recalling that in the coherent case
\begin{align*}
	\capacity_{\mathrm{coh}}(\snr)=\min\{\txant,\rxant\}\log(\snr) + \landauO(1),\quad \snr\to\infty
\end{align*}
one sees that this loss is pronounced when the channel's coherence time $\cohtime$ is small.
The capacity expression~\eqref{eq:zheng_tse_v1} also implies that, for a given coherence time \cohtime and number of receive antennas $\rxant$, the capacity \emph{pre-log} (i.e., the asymptotic ratio between the capacity in \fref{eq:zheng_tse_v1} and $\log(\snr)$ as $\snr\to\infty$) is maximized by using $\txant=\min\{\rxant,\floor{\cohtime/2}\}$ transmit antennas.\footnote{\label{footnote:M_larger_T}More generally, for fixed $\cohtime$ and $\rxant$, and for arbitrary SNR, the capacity for $\txant > \cohtime$ is equal to the capacity for $\txant = \cohtime$~\cite[Thm.~1]{marzetta99-01a}.}
\label{page:footnote}


When $\cohtime\geq \txant +\rxant$  (channel's coherence time larger or equal to the total number of antennas) the high-SNR expression~\eqref{eq:zheng_tse_v1} can be tightened as follows~\cite[Sec. IV.B]{zheng02-02a}:
\begin{align}\label{eq:zheng_tse_v2}
	C(\snr)=\txantopt\left(1-\frac{\txantopt}{\cohtime}\right)\log (\snr) +\const +\landauo(1), \quad \snr \to \infty.
\end{align}
Here,~\const, which is given in~\cite[Eq.~(24)]{zheng02-02a}, depends on \cohtime, \txant, and \rxant but not on \snr, and $\landauo(1)\to 0$ as $\snr \to \infty$.
Differently from~\eqref{eq:zheng_tse_v1}, the high-SNR  expression~\eqref{eq:zheng_tse_v2} describes capacity accurately already at moderate SNR values~\cite{takeuchi11-11a},
 because it captures the first two terms in the asymptotic expansion of $C(\snr)$ for $\snr\to\infty$.
The key element exploited in~\cite{zheng02-02a} to establish~\eqref{eq:zheng_tse_v2} is the optimality of \emph{isotropically distributed unitary} input signals~\cite[Sec.~A.2]{marzetta99-01a} at high SNR. The isotropic unitary input distribution is often referred to as \emph{unitary space-time modulation} (USTM)~\cite{hochwald00-09a,hassibi02-06a,ashikhmin10-11a}.
Capacity-approaching coding schemes that are based on USTM and do not require the explicit estimation of the fading channel have been recently proposed in~\cite{ashikhmin10-11a}.

In this paper, we focus on the case $\cohtime<\txant+\rxant$ (channel's coherence time smaller than the total number of antennas), which is of interest for point-to-point communication systems using large antenna arrays.
The use of large antenna arrays in MIMO systems (\emph{large-MIMO} systems) has been recently advocated to reduce energy consumption in wireless networks, to combat the effect of small-scale fading, and to release multi-user MIMO gains with limited co-operation among base stations and low complexity channel estimation algorithms~\cite{marzetta10-11a,ngo11-12a,rusek11-a}.

\paragraph*{Contributions} 
	\label{par:contributions}

We prove that in the large-MIMO setting where $\cohtime<\txant+\rxant$, USTM is not capacity-achieving at high SNR.
The capacity-achieving input signal turns out to consist of the product of a unitary isotropically distributed random matrix and a diagonal matrix whose nonzero entries are distributed as the square-root of the eigenvalues of a Beta-distributed random matrix of appropriate size.
Utilizing this input distribution, which we refer to as \emph{Beta-variate space-time modulation} (BSTM), we extend~\eqref{eq:zheng_tse_v2} to the case $\cohtime<\txant+\rxant$.
We show that using BSTM instead of USTM yields a rate gain of about $13\%$  when SNR is $30\dB$ and $\rxant\gg\cohtime$. Note that our result holds for all $\cohtime$, $\txant$, and $\rxant$ values satisfying $1<\cohtime<\txant+\rxant$. In other words, differently from most of the literature on large-MIMO systems, our analysis is not asymptotic in the number of antennas.
%

Our proof technique exploits the geometric structure in the MIMO block-fading channel input-output relation first observed in~\cite{zheng02-02a}.
The set of tools used to establish our main result is, however, different from the one used in~\cite{zheng02-02a}.
In particular, differently from~\cite{zheng02-02a}, our proof is based on the duality approach~\cite{lapidoth03-10a}, and on a novel closed-form characterization of the probability density function (pdf) of the MIMO block-fading channel output, which generalizes a previous result obtained in~\cite{hassibi02-06a}.
These two tools allow us to simplify the derivation of \eqref{eq:zheng_tse_v2} for the case $\cohtime \geq \txant+\rxant$ compared to the derivation provided in~\cite{zheng02-02a}, and to generalize~\eqref{eq:zheng_tse_v2} to the large-MIMO setting $\cohtime<\txant+\rxant$.

\paragraph*{Notation} 
\label{par:notation}
Uppercase boldface letters denote matrices and lowercase boldface letters
designate vectors.
Uppercase sans-serif letters (e.g., $\mathsf{Q}$) denote probability distributions, while lowercase sans-serif letters  (e.g., \outpdf) are reserved for pdfs.
The superscripts~$\tp{}$ and~$\herm{}$ stand for
transposition and Hermitian transposition,
respectively.
We denote the identity matrix of dimension~$\txant \times \txant$ by~$\matI_{\txant}$; $\diag\{\veca\}$ is the diagonal square matrix whose main diagonal contains the entries of the vector~\veca, and $\eig{q}{\matA}$ stands for the $q$th largest eigenvalue of the Hermitian positive-semidefinite matrix~\matA.
For a random matrix~$\matX$ with probability distribution~$\inpdist$, we
write~$\matX\distas\inpdist$.
We denote expectation by~$\Ex{}{\cdot}$, and use the
notation~$\Ex{\matX}{\cdot}$ or $\Ex{\inpdist}{\cdot}$ to stress that expectation is taken with
respect to $\matX\distas \inpdist$.
We write~$\relent{\outdist(\cdot)}{\chosoutdist(\cdot)}$ for the
relative entropy between the probability distributions~$\outdist$
and~$\chosoutdist$.
Furthermore, $\jpg(\veczero,\matSigma)$ stands for the distribution of a
circularly-symmetric complex Gaussian random vector with covariance
matrix~$\matSigma$.
For two functions~$f(x)$ and~$g(x)$, the
notation~$f(x) = \landauO(g(x))$, $x\to \infty$, means that
$\lim \sup_{x\to\infty}\bigl|f(x)/g(x)\bigr|<\infty$, and
$f(x) = \landauo(g(x))$, $x\to \infty$, means that $\lim_{x\to\infty}\bigl|f(x)/g(x)\bigr|=0$.
Finally, $\log(\cdot)$ indicates the natural logarithm, $\Gamma(\cdot)$ denotes the Gamma function~\cite[Eq.~(6.1.1)]{abramowitz72-a}, and $\Gamma_{m}(a)$ designates the \emph{complex} multivariate Gamma function~\cite[Eq.~(44)]{marques08-03a}
\begin{IEEEeqnarray}{c}
\label{eq:def-complex-multi-gamma}
\Gamma_{m}(a) =\pi^{m(m-1)/2}\prod_{k=1}^m\Gamma(a-k+1).
\end{IEEEeqnarray}
\section{System Model and Known Results} 
\label{sec:system_model}
\subsection{System Model}
\label{sec:system_model_subsection}

We consider a point-to-point Rayleigh block-fading MIMO channel with \txant transmit antennas, \rxant receive antennas, and channel's coherence time~$\cohtime>1$.
The channel input-output relation within a coherence interval can be compactly written in matrix notation as follows~\cite{zheng02-02a,hochwald00-03a,hassibi02-06a}:
\begin{align}\label{eq:channel_IO}
	\matY=\sqrt{{\snr}/{\txant
	}}\cdot\matX\matH+\matW.
\end{align}
Here,~$\matX=[\vecx_1\, \cdots\, \vecx_{\txant}] \in \complexset^{\cohtime \times \txant}$ contains the signal transmitted from the \txant antennas within the coherence interval, $\matH \in \complexset^{\txant \times \rxant}$ is the channel's propagation matrix,~$\matW \in \complexset^{\cohtime \times \rxant}$ is the additive noise, and $\matY \in \complexset^{\cohtime \times \rxant}$ contains the signal received at the \rxant antennas within the coherence interval.
We will assume throughout the paper that $\txant\leq\min\{\rxant,\floor{\cohtime/2}\}$. \label{page:assumption-on-M}
The random matrices \matH and \matW are independent of each other and have independent and identically distributed (\iid) $\jpg(0,1)$ entries.
We consider the noncoherent setting where neither the transmitter nor the receiver have \emph{a priori} knowledge of the realizations of \matH and \matW, but both know their statistics perfectly.

We assume that \matH and \matW take on independent realizations over successive coherence intervals.
Under this \emph{block-memoryless} assumption, the ergodic capacity of the channel in~\eqref{eq:channel_IO}  is given by
\begin{align}\label{eq:definition_capacity}
	  C(\snr) =\frac{1}{\cohtime} \sup_{\inpdist}\mi(\matX;\matY).
\end{align}
Here, $\mi(\matX;\matY)$ denotes the mutual information~\cite[Sec.~8.5]{cover06-a} between the input matrix~\matX and the output matrix~\matY, and the supremum is over all probability distributions~\inpdist on~\matX that satisfy the average-power constraint
\begin{equation}
\label{eq:average-power-constraint}
\Ex{}{\tr\{\matX\herm{\matX}\}} \leq \cohtime\txant.
\end{equation}
Since the variance of the entries of \matH and \matW is normalized to one,~\snr in~\eqref{eq:channel_IO} can be interpreted as the SNR at each receive antenna.

\begin{table}
\extrarowheight=3pt
	\begin{center}
	\begin{tabular}{|c|c|}
		\hline
		   Parameter & Definition \\ \hline \hline
		   $\maxparam$ & $   \max\{\rxant,\cohtime-\txant\}$   \\ \hline
           $ \minparam$ &  $   \min\{\rxant,\cohtime-\txant\}$  \\ \hline
		   $\maxbound$ & $\max\{\rxant,\cohtime\}$ \\  \hline
           $\minbound$ & $\min\{\rxant,\cohtime\}$  \\ \hline
	\end{tabular}
	\end{center}
	\caption{Four parameters related to the channel's coherence time $\cohtime$, the number of transmit antennas $\txant$, and the number of receive antennas $\rxant$.}
	\label{tab:simple_functions}
\vspace{-1cm}
\end{table}
Throughout the paper, we will often make use of four parameters ($\minparam$, $\maxparam$, $\minbound$, $\maxbound$) related to the coherence time $\cohtime$, the number of transmit antennas $\txant$, and the number of receiver antennas $\rxant$.
These parameters are listed in~\fref{tab:simple_functions} for future reference.

%

\subsection{Properties of the Capacity-Achieving Input Distribution} 
\label{sec:properties_of_the_capacity_achieving_distribution}

Even if  no closed-form expression is available to date for $\capacity(\snr)$, the structure of the capacity-achieving input distribution is partially known.
We next review two properties of the capacity-achieving input distribution that will reveal useful for our analysis.
\begin{lem}[\!{\cite[Thm.~2]{marzetta99-01a}}]\label{lem:structure_cap_ach_distr}
	The capacity-achieving input matrix \matX is the product of a $\cohtime \times \txant$ isotropically distributed unitary matrix \matPhi and an independent $\txant \times \txant$ nonnegative diagonal matrix $\matD=\diag\{\tp{[d_1\,\cdots\, d_\txant]}\}$.
\end{lem}

For the case $\cohtime\geq \txant +\rxant$, taking \matD deterministic with diagonal entries equal to $\sqrt{\cohtime}$ turns out to be optimal at high SNR.
In this case, the resulting input matrix \matX is a scaled isotropically distributed unitary matrix.
This input distribution, which is known as USTM~\cite{hochwald00-09a,hassibi02-06a,ashikhmin10-11a}, is the one used in~\cite{zheng02-02a} to establish~\eqref{eq:zheng_tse_v2}.

When $\cohtime < \txant + \rxant$, USTM is not optimal at high SNR, as we shall illustrate in~\fref{sec:main_result}.
Nevertheless, the optimal distribution of $\matX=\matPhi\matD$ shares the following property with USTM: the probability distribution induced on $\sqrt{\snr}d_m=\sqrt{\snr}\vecnorm{\vecx_{m}}$, $m=1,\dots,\txant$, by the capacity-achieving input distribution \emph{escapes to infinity}~\cite[Def.~4.11]{lapidoth03-10a} as $\snr\to \infty$. Namely, it allocates vanishing probability to every interval of the form $\bigl[0,\sqrt{\snr_0}\bigr]$ with $\snr_0>0$.
This property is formalized in the following lemma:
\begin{lem}
\label{lem:escape-to-infty}
Fix an arbitrary $\snr_0>0$ and let
\begin{IEEEeqnarray}{rCl}
\label{eq:setK}	\mathcal{K}(\snr_0)\define \Big\{ \matA=[\veca_1\,\cdots\,&&\veca_\txant]\in \complexset^{\cohtime\times\txant}\sothat\notag\\
&& \min\limits_{m=1,\ldots,\txant}\{\snr\|\veca_m\|^2\} < \snr_0 \Big\}.
\end{IEEEeqnarray}
Let $\{\inpdist^{(\snr)}, \snr > 0\}$ be a family of input distributions (parametrized with respect to the SNR \snr) satisfying (\ref{eq:average-power-constraint}) and the following additional property
\begin{equation*}
\lim\limits_{\snr\rightarrow \infty} \frac{\mi(\matX;\matY) }{C(\snr)} = 1, \quad \matX \distas \inpdist^{(\snr)}.
\end{equation*}
Then,
$\lim\nolimits_{\snr\rightarrow\infty} \mathbb{P}\lefto(\matX \in \mathcal{K}(\snr_0),\,\matX\distas\inpdist^{(\snr)}\right) =0$.
%
\end{lem}
\begin{IEEEproof}
   The proof follows along the same lines of the proofs of ~\cite[Thm.~8]{lapidoth06-02a} and~\cite[Lem~8]{zheng02-02a}.
\end{IEEEproof}	

An important consequence of the escape-to-infinity property of the capacity-achieving input distribution is that the asymptotic behavior of $\capacity(\snr)$ as $\snr \to \infty$ does not change if we constrain the probability distribution of $\sqrt{\snr}d_m$ ($\allo{m}{\txant}$) to be supported outside the interval $[0,\sqrt{\snr_0}]$, $\snr_0>0$.
More precisely, we have the following result.
\begin{lem}\label{lem:outside_ball}
	Fix an arbitrary $\snr_0>0$ and let $\setK(\snr_0)$ as in~\eqref{eq:setK}.
	Denote by $C_{\setK}(\snr)$ the capacity of the channel~\eqref{eq:channel_IO} when the input \matX is subject to the average-power constraint~\eqref{eq:average-power-constraint} and to the additional constraint that $\matX \notin \setK(\snr_0)$ with probability~1 (\wpone).
	Then,
	%
		$C(\snr)=C_{\setK}(\snr) + \landauo(1), \, \snr \to \infty$.
	%
\end{lem}
\begin{IEEEproof}
   The proof follows from~\cite[Thm.~4.12]{lapidoth03-10a}.
\end{IEEEproof}


\section{Capacity in the High-SNR Regime} 
\label{sec:main_result}
\subsection{Asymptotic Characterization of Capacity}
\label{sec:a_complete_asymptotic_characterization_of_capacity}
The main result of this paper is~\fref{thm:asymptotics} below, which provides a high-SNR characterization of $\capacity(\snr)$ that generalizes~\eqref{eq:zheng_tse_v2}, in that it holds also in the large-MIMO setting $\cohtime<\txant+\rxant$.\footnote{\label{footnote:MIMO_uplink_receiver}Because of the constraint $\txant \leq \min\{\rxant,\lfloor{\cohtime}/{2}\rfloor\}$, \emph{large-MIMO setting} in this paper indicates a point-to-point MIMO uplink with a large antenna array at the receiver.}
\begin{thm}\label{thm:asymptotics}
  The capacity~$\capacity(\snr)$ of the Rayleigh block-fading MIMO channel~\eqref{eq:channel_IO} with \rxant receive antennas, coherence time~\cohtime, and~$\txant\leq \min\{\rxant, \floor{\cohtime/2} \}$ transmit antennas is given by
\begin{IEEEeqnarray}{rCl}\label{eq:asymptotics}
	C(\snr) &=&  \txant\left(1- \frac{\txant}{\cohtime}\right)\log (\snr) +\const+ o(1),\quad \snr\rightarrow \infty
\end{IEEEeqnarray}
where
\begin{IEEEeqnarray}{rCl}\label{eq:const_value_general}
	\const&\define&\frac{1}{\cohtime}\log\lefto( \frac{\Gamma_{\txant}(\txant)\Gamma_{\txant}(\minparam)}{\Gamma_\txant(\rxant)\Gamma_\txant(\cohtime)}\right) + \txant\left(1-\frac{\txant}{\cohtime}\right)\log\lefto(\frac{\cohtime}{\txant}\right)\nonumber\\
	 &&+\:\frac{\txant \minparam}{\cohtime} \log \lefto(\frac{\rxant}{\minparam}\right) + \frac{\maxparam}{\cohtime
	}\Bigl(\Ex{}{\logdet{\matH\herm{\matH}}}- \txant \Bigr).
\end{IEEEeqnarray}
Here, \minparam and \maxparam are defined in \fref{tab:simple_functions}, and
\begin{IEEEeqnarray}{rCl}
\label{eq:exp_log_det_and_euler_digamma}
\Ex{}{\logdet{\matH\herm{\matH}}} &=&
\sum_{i=1}^{\txant} \psi(\rxant-i+1) \notag\\
&=&-\txant\gamma +\sum\limits_{i=1}^{\txant}\sum\limits_{k=1}^{\rxant-i}\frac{1}{k}
\end{IEEEeqnarray}
where~$\psi(\cdot)$ denotes Euler's digamma function~\cite[Eq.~(6.3.1)]{abramowitz72-a} and~$\gamma\approx 0.577$ is Euler's constant.
\end{thm}

\begin{IEEEproof}
 See \fref{sec:proof_of_thm:asymptotics}. A sketch of the proof for the single-input multiple-output case, which is simpler to analyze than the MIMO case, is given in~\cite{yang12-07a}.
\end{IEEEproof}

In~\fref{sec:rate_achievable_with_ustm} below we compare $C(\snr)$ in~\eqref{eq:asymptotics}  with the capacity lower bound obtained using USTM.
The input distribution that achieves~\eqref{eq:asymptotics} is described in~\fref{sec:the_capacity_achieving_distribution_at_high_snr}.
Numerical results illustrating the lack of tightness of the USTM-based capacity lower bound  in the large-MIMO setting are provided in \fref{sec:gain_of_BSTM}.
%

\subsection{Rates Achievable with USTM} 
\label{sec:rate_achievable_with_ustm}

	For the case $\cohtime\geq \txant + \rxant$, the high-SNR capacity expression~\eqref{eq:asymptotics}  coincides with the one reported in \cite[Sec. IV.B]{zheng02-02a}.\footnote{The expression for \const given in~\cite[Eq.~(24)]{zheng02-02a}  contains a typo: the argument of the logarithm in the second addend should be divided by $\txant$ as one can verify by comparing~\cite[Eq.~(24)]{zheng02-02a} with the result given in~\cite[Thm.~9]{zheng02-02a} for the case $\txant=\rxant$.}
	In this case, USTM, i.e., $\matX=\sqrt{\cohtime}\matPhi$, with \matPhi unitary and isotropically distributed, achieves~\eqref{eq:asymptotics}.
	When $\cohtime< \txant + \rxant$, the novel capacity characterization provided in~\fref{thm:asymptotics} implies that USTM is not capacity-achieving at high SNR, as formalized in the following corollary.

\begin{cor}\label{cor:const_mod_input}
	The rate achievable using USTM over the Rayleigh block-fading MIMO channel~\eqref{eq:channel_IO} with \rxant receive antennas, coherence time~\cohtime, and~$\txant\leq \min\{\rxant, \floor{\cohtime/2} \}$ transmit antennas is
	\begin{align}\label{eq:asymptotics_const_mod}
		\custm(\snr) &=  \txant\left(1- \frac{\txant}{\cohtime}\right)\log (\snr) +\constcm+ o(1),\, \snr\rightarrow \infty
	\end{align}
	where
	\begin{IEEEeqnarray*}{rCl}
		\constcm &\define& \frac{1}{\cohtime}\log\lefto( \frac{\Gamma_{\txant}(\txant)}{\Gamma_\txant(\cohtime)}\right) + \txant\left(1-\frac{\txant}{\cohtime}\right)\log\lefto(\frac{\cohtime}{e\txant }\right) \\&&+ \left(1-\frac{\txant}{\cohtime}\right) \Ex{}{\log\det(\matH\herm{\matH})}.
	\end{IEEEeqnarray*}
\end{cor}

Note that $\constcm=\const$ when $\cohtime\geq \txant + \rxant$; however, $\constcm< \const$ when $\cohtime < \txant + \rxant$.
\begin{IEEEproof}
	The proof follows by repeating the same steps as in~\fref{sec:lower_bound} after having replaced the capacity-achieving input distribution (to be described in~\fref{sec:the_capacity_achieving_distribution_at_high_snr}) with USTM.
\end{IEEEproof}

\subsection{The Capacity-Achieving Input Distribution at High SNR} 
\label{sec:the_capacity_achieving_distribution_at_high_snr}

\subsubsection{Matrix-variate distributions} 
\label{sec:complex_wishart_and_beta_matrix_variate_distributions}
To describe the input probability distribution that achieves~\eqref{eq:asymptotics}, we shall need the following preliminary results from multivariate statistics.
\begin{dfn}
\label{dfn:Wishart}
	An $m\times m$ random matrix \matA is said to have a \emph{complex Wishart distribution} with $n>0$ degrees of freedom and covariance matrix \matSigma if $\matA=\matB\herm{\matB}$, where the columns of the $m \times n$ matrix \matB are independent and $\jpg(\veczero,\matSigma)$-distributed.
	In this case, we shall write $\matA\distas\wishart_{m}(n,\matSigma)$.
\end{dfn}

Note that when $m>n$, the matrix \matA is singular and, hence, does not admit a pdf.
In this case, the probability distribution of \matA is sometimes referred to as \emph{pseudo-Wishart} or \emph{singular Wishart}.
%
%

\begin{dfn}
\label{dfn:complex-matrix-beta}
	An $m \times m$ random matrix \matC is said to have a \emph{complex  matrix-variate Beta distribution} of parameters $p>0$ and $n>0$ if \matC can be written as
	%
	$	\matC=\bigl(\herm{\matT}\bigr)^{-1}\matA\matT^{-1}$,
	%
	where $\matA\distas\wishart_m(p,\matSigma)$ and $\matB\distas \wishart_m(n,\matSigma)$ are independent, and $\matA+\matB=\herm{\matT}\matT$, with \matT upper-triangular  with positive diagonal elements~\cite[p.~406]{horn85a}.
	In this case, we shall write $\matC\distas\betadist_m(p,n)$.
\end{dfn}

For the case when $n<m$ or $p<m$,  the probability distribution of $\matC$ is usually referred to as singular complex matrix-variate Beta distribution because it involves singular Wishart distributions.
In the next lemma, we state two properties of the complex matrix-variate Beta  distribution that will be used in the proof of \fref{thm:asymptotics}.

\begin{lem}\label{lem:properties_beta}
   Let $\matC\distas\betadist_m(p,n)$ with $p\geq m>0$ and $n>0$.
The following properties hold:
\begin{enumerate}
	\item \matC is unitarily invariant~\cite[Def.~2.6]{tulino04a}, i.e., $\matC\distas \matU\matC\herm{\matU}$ for every $m\times m$ unitary matrix \matU independent of \matC.
	\item The joint pdf of the ordered eigenvalues $\lambda_1\geq\dots\geq \lambda_m$ of \matC takes on two different forms according to the value of $n$.
	If $n\geq m$, then $1>\lambda_1>\dots> \lambda_m>0$ \wpone, and the joint pdf of $\lambda_1,\dots, \lambda_m$ is given by
	\begin{IEEEeqnarray}{rCl}
	   \IEEEeqnarraymulticol{3}{l}{\mathsf{f}_{\lambda_1,\ldots,\lambda_m}(a_1,\ldots,a_m)}\notag \\
\quad &=& \frac{\pi^{m(m-1)}}{\Gamma_m(m)}\cdot\frac{\Gamma_m(p+n)}{\Gamma_m(p)\Gamma_m(n)}\notag\\
&& \cdot\prod\limits_{i=1}^{m}a_i^{p-m}(1-a_i)^{n-m}\cdot\biprod{i<j}{m}(a_i-a_j)^2.\label{eq:pdf_beta_nonsingular}
	\end{IEEEeqnarray}
	If $0<n<m$, then $\lambda_1=\cdots=\lambda_{m-n}=1$ \wpone, and $1>\lambda_{m-n+1}>\cdots>\lambda_m>0$ \wpone. Moreover, the joint pdf of $\lambda_{m-n+1},\dots,\lambda_m$ is given by
	\begin{IEEEeqnarray}{rCl}
		\IEEEeqnarraymulticol{3}{l}{\mathsf{f}_{\lambda_{m-n+1},\ldots,\lambda_m}(a_{m-n+1},\ldots,a_{m})}\nonumber\\
\,\,&=&\frac{\pi^{n(n-1)}}{\Gamma_{n}(n)}\cdot\frac{\Gamma_{n}(p+n)}{\Gamma_{n}(m)\Gamma_{n}(p+n-m)}\notag\\
&&\cdot\!\!\prod\limits_{i=m-n+1}^{m}\!\!\!\!\!a_i^{p-m}(1-a_i)^{m-n}\cdot\!\!\!\!\prod\limits_{m-n<i<j}^{m}\!\!\!(a_i-a_j)^2.
\label{eq:pdf_beta_singular}\IEEEeqnarraynumspace
\end{IEEEeqnarray}
\end{enumerate}
\end{lem}
\begin{IEEEproof}
   Part 1 and~\eqref{eq:pdf_beta_nonsingular} in part~2 follow by extending to the complex case~\cite[Lem.~3.11]{mitra70-03a} and~\cite[Thm.~3.3.4]{muirhead05}, respectively; to prove~\eqref{eq:pdf_beta_singular} it is sufficient to note that $\widetilde{\matC}=(\matI_m-\matC)\distas\betadist_{m}(n,p)$ (see~\cite[Def.~3.3.2]{muirhead05}), that $\widetilde{\matC}$ has rank $n<m$, and that its $n$ nonzero eigenvalues are distributed as the eigenvalues of a $\betadist_n(m,p+n-m)$-distributed random matrix.
\end{IEEEproof}

We shall also need the following result relating Wishart-distributed and Beta-distributed matrices.
\begin{lem}\label{lem:wishart_and_beta}
	Let $\matS\distas\wishart_m(p+n,\matSigma)$ with $m>0$, $n>0$, and $p\geq m$.
	Furthermore, let $\matC\distas\betadist_m(p,n)$ be independent of $\matS$.
   	Finally, put $\matS=\herm{\matT}\matT$, where $\matT$ is upper-triangular with positive diagonal elements.
	Then,
%
	$\matA=\herm{\matT}\matC\matT\distas\wishart_{m}(p,\matSigma)$.
%
\end{lem}
\begin{IEEEproof}
	The lemma follows from a generalization to the complex case of~\cite[Thm.~3.3.1]{muirhead05} for the nonsingular case $n\geq m$, and of~\cite[Thm.~1]{uhlig94-a} for the singular case $0<n<m$.
\end{IEEEproof}

Note that \fref{lem:properties_beta} (part 1) implies that the eigenvalues of \matA and of $\matC\matS$ in \fref{lem:wishart_and_beta} have the same distribution.

\subsubsection{The Optimal Input Distribution} 
\label{sec:the_optimal_distribution_on_matd}

We are now ready to describe the input distribution that achieves~\eqref{eq:asymptotics}.
This distribution takes on two different forms according to the relation between $\cohtime, \txant$ and \rxant.
Specifically, one should take $\matX=\matPhi\matD$ where $\matPhi$ is unitary and isotropically distributed, and
$\matD=\sqrt{{\cohtime\rxant}/{\minparam}}\cdot\auxdiagmat$
with \minparam  defined in \fref{tab:simple_functions}, and with $\widetilde{\matD}$ being a diagonal matrix whose \emph{ordered} positive entries $\{\auxdiagent_1,\dots,\auxdiagent_{\txant}\}$ are distributed as follows:
\paragraph{Case $\cohtime<\txant+\rxant$} 
\label{par:_cohtimeleqrxant_}
The squared nonzero entries $\{\auxdiagent_1^2,\dots,\auxdiagent^2_{\txant}\}$ of \auxdiagmat have the same joint pdf as the ordered eigenvalues of a positive-definite $\txant \times \txant$ random matrix $\matZ\distas\betadist_{\txant}(\cohtime-\txant,\txant+\rxant-\cohtime)$.
The resulting pdf of $\{\auxdiagent_1^2,\dots,\auxdiagent^2_{\txant}\}$ is obtained by setting $p=\cohtime-\txant$ and $n=\txant+\rxant-\cohtime$ in~\eqref{eq:pdf_beta_nonsingular} if $\cohtime\leq\rxant$, and in~\eqref{eq:pdf_beta_singular} if $\rxant<\cohtime<\txant+\rxant$.
%
%
\paragraph{Case $\cohtime\geq \txant +\rxant$} 
\label{par:case_cohtimegeq_txant_rxant_}
The nonzero entries $\{\auxdiagent_1,\dots,\auxdiagent_{\txant}\}$ of \auxdiagmat should be taken so that
%
   $\auxdiagent_1=\dots=\auxdiagent_{\txant}= 1$
%
\wpone.
This results in the USTM distribution used in~\cite{zheng02-02a}.

In the remainder of the paper, we shall denote by~$\distoptD$ the probability distribution of $\matD=\sqrt{{\cohtime\rxant}/{\minparam}}\cdot\auxdiagmat$ we have just introduced.
Furthermore, we shall refer to the probability distribution of $\matX=\matPhi\matD$ resulting by choosing \matPhi unitary and isotropically distributed and $\matD\distas\distoptD$ as BSTM.
Note that BSTM reduces to USTM when $\cohtime\geq \txant+\rxant$.

 As shown in~\cite[p.~369]{zheng02-02a}, USTM is optimal for the case $\cohtime \geq \txant +\rxant$ because it maximizes
\begin{IEEEeqnarray}{c}
\label{eq:terms-from-mutual-info}
   \difent(\matU\matD\matH)+(\cohtime-\txant-\rxant)\Ex{}{\logdet{\matD^2}}
\end{IEEEeqnarray}
where $\matU \in \complexset^{\txant \times \txant}$ is an isotropically distributed unitary matrix independent of both $\matD$ and $\matH$, and $\difent(\cdot)$ denotes the differential entropy.
In fact, the average-power constraint~\eqref{eq:average-power-constraint} implies that
\begin{IEEEeqnarray*}{c}
\difent(\matU\matD\matH) \leq \txant \rxant \log\lefto(\pi e \cohtime\right);\,\,\,\,\,\Ex{}{\log\det(\matD^2)}\leq \txant \log(\cohtime)
\end{IEEEeqnarray*}
and under USTM, which yields $\matD=\sqrt{\cohtime}\cdot\matI_\txant$, both inequalities hold with equality.

In the large-MIMO setting $\cohtime <\txant + \rxant$, however, the second term in~\eqref{eq:terms-from-mutual-info} turns negative and USTM does no longer maximize~\eqref{eq:terms-from-mutual-info}. As we shall now illustrate, the maximizing distribution of $\matD$ turns out to be~$\distoptD$, which results in BSTM. Through algebraic manipulations similar to the ones leading to \eqref{eq:lb-difent-Y-2} and \eqref{eq:lb-difent-G} in~\fref{sec:proof_of_thm:asymptotics}, it is possible to show that
 \begin{IEEEeqnarray*}{rCl}
  \difent(\matU\matD\matH)+(\cohtime-\txant-\rxant)\Ex{}{\logdet{\matD^2}} = \difent (\matG) + k.
 \end{IEEEeqnarray*}
Here, $k$ is a constant that does not dependent on $\matD$, and $\matG \in \complexset^{\txant \times (\cohtime -\txant)}$ is a random matrix with singular values jointly distributed as the singular values of $\matD\matH$, and with isotropically distributed singular vectors.
Lemma~\ref{lem:BSTM_and_noiseless_output} below implies that the choice $\matD\sim\distoptD$ induces a matrix $\matG$ that is Gaussian with i.i.d. $\jpg(0,{\cohtime\rxant}/{(\cohtime-\txant)})$ entries. But a Gaussian $\matG$ with i.i.d. entries maximizes $\difent(\matG)$, and, hence,~\eqref{eq:terms-from-mutual-info}.
\begin{lem}\label{lem:BSTM_and_noiseless_output}
	Let $\matD\distas\distoptD$ and let $\matH$ be an independent $\txant \times \rxant$ random matrix with \iid $\jpg(0,1)$ entries.
	The singular values of $\matD\matH$ are distributed as the singular values of an $\txant \times \minparam$ matrix \matG with \iid $\jpg(0,\cohtime\rxant/\minparam)$ entries.
\end{lem}
\begin{IEEEproof}
	For the case $\cohtime\geq \txant+\rxant$, we have that $\minparam=\rxant$ and, hence, $\matD=\sqrt{\cohtime}\cdot\matI_{\txant}$.
	Consequently, $\matD\matH=(\sqrt{\cohtime}\cdot\matH)\distas\matG$, from which the statement in the lemma follows.

	For the case $\cohtime<\txant+\rxant$ (and, hence, $\minparam=\cohtime-\txant$) we shall proceed as follows.
	Let $\matD=\sqrt{\cohtime\rxant/(\cohtime-\txant)}\cdot\auxdiagmat$, and let $\matU$ be an $\txant \times \txant$ unitary and isotropically distributed random matrix independent of $\auxdiagmat$ and \matH.
	Since $\matH\herm{\matH}$ is unitary invariant, we have that $\matH\herm{\matH}\distas\herm{\matU}\matH\herm{\matH}\matU$, and hence $\auxdiagmat\matH\herm{\matH}\auxdiagmat\distas \auxdiagmat\herm{\matU}\matH\herm{\matH}\matU\auxdiagmat$.
	Now note that $\auxdiagmat\herm{\matU}\matH\herm{\matH}\matU\auxdiagmat$ and $\matU\auxdiagmat^2\herm{\matU}\matH\herm{\matH}$ have the same eigenvalues;
	furthermore,  $\matU\auxdiagmat^2\herm{\matU}\distas \betadist_{\txant}(\cohtime-\txant,\txant+\rxant-\cohtime)$, which follows from~\fref{lem:properties_beta} (part 1), and from \cite[Lem.~2.6]{tulino04a}; finally,~$\matH\herm{\matH}\distas\wishart_\txant(\rxant,\matI_\txant)$.
	Hence, by Lemmas \ref{lem:properties_beta} and \ref{lem:wishart_and_beta} the eigenvalues of $\matU\auxdiagmat^2\herm{\matU}\matH\herm{\matH}$---and consequently also the eigenvalues of  $\auxdiagmat\matH\herm{\matH}\auxdiagmat$---have the same distribution as the eigenvalues of a $\wishart_{\txant}(\cohtime-\txant,\matI_{\txant})$-distributed random matrix.
\end{IEEEproof}

\subsection{Gain of BSTM over USTM} 
\label{sec:gain_of_BSTM}

	The use of USTM is motivated by several practical considerations~\cite{hochwald00-03a,hassibi02-06a,ashikhmin10-11a}.
	Is it then worth to replace USTM by the capacity-achieving BSTM in the large-MIMO setting?
	 In this section, we shall investigate the rate gain that results from the use of BSTM instead of USTM.

\paragraph*{Asymptotic Analysis} 
	\label{par:asymptotic_analysis}

	In \fref{cor:gain}  below we show that the rate gain resulting from the use of BSTM instead of USTM grows logarithmically in the number of receive antennas.
	\begin{cor}\label{cor:gain}
		Let $\cohtime$ and $\txant\leq\floor{\cohtime/2}$ be fixed.
		Then
		\begin{IEEEeqnarray}{c}\label{eq:gain}
		\lim\limits_{\rxant \rightarrow \infty} \lim\limits_{\snr\rightarrow \infty}\!\! \left(C(\snr) -\custm(\snr) - \frac{\txant^2}{2\cohtime} \log (\rxant) \right) = c_{\txant,\cohtime}\IEEEeqnarraynumspace
		\end{IEEEeqnarray}
		where~$C(\snr)$ and $\custm(\snr)$ are given in~\eqref{eq:asymptotics} and~\eqref{eq:asymptotics_const_mod}, respectively, and
		\begin{IEEEeqnarray*}{rCl}
		c_{\txant,\cohtime} &\define& \frac{1}{\cohtime}\log\Bigl(\Gamma_\txant(\cohtime - \txant)\Bigr) + \frac{\txant(\cohtime - \txant)}{\cohtime}\log\lefto(\frac{e}{\cohtime-\txant}\right)
		 \\
        &&-\, \frac{\txant}{2 \cohtime}\Bigl[\txant\log(\pi e)+\log( 2)\Bigr].
		\end{IEEEeqnarray*}
	\end{cor}
	\begin{IEEEproof}
		As we are interested in the limit $\rxant\to \infty$,  we shall assume without loss of generality that $\minparam=\cohtime-\txant$ and $\maxparam=\rxant$.
		 Since the first term in the high-SNR expansion of $C(\snr)$ and $\custm(\snr)$ is the same,
		\begin{align*}
			  \lim\limits_{\snr\rightarrow \infty} \Bigl(C(\snr) -\custm(\snr)\Bigr)&=\const-\constcm = c_0 + c_\rxant
		\end{align*}
		where $c_0$ and $c_\rxant$ are defined as follows:
		\begin{IEEEeqnarray}{rCl}
\cohtime\cdot c_0&=& \log\Bigl( \Gamma_{\txant}(\cohtime-\txant)\Bigr) + \txant(\cohtime-\txant)\log\lefto( \frac{e}{\cohtime-\txant}\right)\notag\\
			\cohtime \cdot c_{\rxant}&=&(\rxant-\cohtime+\txant)\Ex{}{\logdet{\matH\herm{\matH}}}-\log \bigl(\Gamma_{\txant}(\rxant)\bigr)\notag\\
            &&-\:\txant\rxant + \txant(\cohtime-\txant)\log (\rxant) \label{eq:part_depending_on_rxant}	    	.
	    \end{IEEEeqnarray}
	    Note that $c_{\rxant}$ is a function of $\rxant$, while $c_0$ is not.
		Consequently, to establish~\eqref{eq:gain} it is sufficient to study the limit $\rxant \to \infty$ of the first two terms on the right-hand side (RHS) of~\eqref{eq:part_depending_on_rxant}.
		For the first term, we use~\eqref{eq:exp_log_det_and_euler_digamma} and the following asymptotic expansion of the Euler's digamma function~\cite[Eq.~(6.3.18)]{abramowitz72-a}:
		%
		   $ \psi(m)=\log (m) - {1}/(2m) + \landauo\lefto({1}/{m}\right), \, m\to \infty $,
		%
		which yields
		\begin{IEEEeqnarray}{rCl}
			\IEEEeqnarraymulticol{3}{l}{(\rxant-\cohtime+\txant)\Ex{}{\logdet{\matH\herm{\matH}}}} \nonumber\\
            \quad&=& (\rxant-\cohtime+\txant) \sum_{i=1}^{\txant} \psi(\rxant-i+1) \nonumber \\
			     &=& -\txant(\cohtime-\txant)\log (\rxant) +\rxant\sum_{i=1}^{\txant}\log(\rxant-i+1) \notag\\
            &&-\: \frac{\txant}{2} +\landauo(1), \quad \rxant \to \infty.\label{eq:first_term_as}
		\end{IEEEeqnarray}
		For the second term on the RHS of~\eqref{eq:part_depending_on_rxant} we proceed as follows:
		\begin{IEEEeqnarray}{rCl}
			\IEEEeqnarraymulticol{3}{l}{\log \Bigl(\Gamma_{\txant}(\rxant)\Bigr)}\notag\\
\quad &=& \frac{\txant(\txant-1)}{2}\log(\pi)+ \sum\limits_{i=1}^{\txant}\log \Bigl((\rxant -i)!\Bigr) \nonumber\\
			&\stackrel{(a)}{=}&\sum\limits_{i=1}^{\txant}\left( (\rxant -i)\log(\rxant -i) +\frac{\log(\rxant-i)}{2} + i \right) - \txant\rxant \nonumber \\ &&+\: \frac{\txant}{2}\log( 2)
			 + \frac{\txant^2}{2} \log(\pi) +o(1),\, \rxant\rightarrow\infty \nonumber\\
			%
            &=& \rxant\left[\sum_{i=1}^\txant \log(\rxant-i) \right] + \frac{\txant^2}{2} \log\left(\frac{\pi e}{\rxant}\right)+\frac{\txant}{2}\log(2e)\notag\\
            &&-\:\txant\rxant+\landauo(1), \, \rxant\to \infty. \label{eq:second_term_as}
		\end{IEEEeqnarray}
		Here,~(a) follows from Stirling's formula $n! = n^n e^{-n}\sqrt{2\pi n}\left(1+o(1)\right), \, n\to \infty$.
		We complete the proof by substituting~\eqref{eq:first_term_as}  and~\eqref{eq:second_term_as} into~\eqref{eq:part_depending_on_rxant}, and using that
		\begin{IEEEeqnarray*}{+rCl+x*}
			\lim_{\rxant \to \infty} \rxant \log\lefto(\frac{\rxant-i+1}{\rxant-i}\right)&= &1. & \IEEEQED
		\end{IEEEeqnarray*}
\let\IEEEQED\relax

	\end{IEEEproof}
	\begin{figure}[!t]
		\centering
			\includegraphics[width=3.5in]{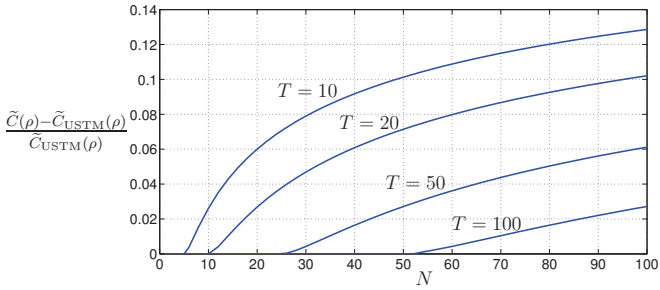}
		      \caption{Rate gain resulting from the use of BSTM instead of USTM as a function of the number of receive antennas \rxant and the channel's coherence time \cohtime; in the figure, $\snr=30\dB$, and $\txant = \min\{\floor{\cohtime/2},\rxant\}$. }
		\label{fig:gain}
	\end{figure}

	\paragraph*{Numerical Results} 
	\label{par:numerical_results}
	   Let~$\approxc(\snr)$ be the high-SNR approximation of $C(\snr)$ obtained by neglecting the $\landauo(1)$ term in~\eqref{eq:asymptotics}.
	   Similarly,~let  $\approxcustm(\snr)$ be the high-SNR approximation of $\custm(\snr)$ obtained by neglecting the $\landauo(1)$ term in~\eqref{eq:asymptotics_const_mod}.
As can be inferred from the results reported in~\cite{zheng02-02a,hassibi02-06a,takeuchi11-11a}, $\approxcustm(\snr)$ is a good approximation for $\custm(\snr)$ at $\snr\gtrsim 30\dB$.
Numerical evidence suggests that the same holds for the pair $\approxc(\snr)$ and $C(\snr)$.
	To illustrate the gain resulting from the use of BSTM instead of USTM for a finite (but large) number of receive antennas, we plot in \fref{fig:gain} the ratio $[\approxc(\snr)-\approxcustm(\snr)]/\approxcustm(\snr)$ for different values of $\cohtime$ and \rxant, when $\snr=30\dB$ and $\txant=\min\{\floor{\cohtime/2},\rxant\}$.
	We observe from~\fref{fig:gain} that the rate gain resulting from the use of BSTM instead of USTM becomes significant when the number of receive antennas $\rxant$ is much larger than the channel's coherence time $\cohtime$.
	For example, when $\rxant=100$ and $\cohtime=10$, the rate gain amounts to $13\%$.
	However, when $\cohtime = \rxant =100$ the rate gain is below $3\%$.
	


\section{Proof of \fref{thm:asymptotics}} 
\label{sec:proof_of_thm:asymptotics}

The proof is effected by exhibiting capacity upper and lower bounds that agree up to a $\landauo(1)$ term.

\subsection{Upper Bound} 
\label{sec:upper_bound}
Fix $\snr_0>0$ and let~$\setK(\snr_0)$ as in~\eqref{eq:setK};
as a consequence of~\fref{lem:outside_ball}, we can restrict---without loss of generality---the supremum in~\eqref{eq:definition_capacity} to  input distributions $\inpdist$ satisfying the constraint $\matX \notin \setK(\snr_0)$ \wpone.
Our capacity upper bound is based on \emph{duality}~\cite{lapidoth03-10a,durisi11-08a}, which is a technique that allows one to obtain tight upper bounds on $I(\matX;\matY)$ by carefully choosing a probability distribution of~\matY.
Specifically, let $\condoutdist$ denote the conditional probability distribution of $\matY$ given $\matX$, and $\outdist$ denote the distribution induced on $\matY$ by $\inpdist$ through~\eqref{eq:channel_IO}. Finally, let $\chosoutdist$ be an arbitrary distribution of $\matY$ with pdf $\chosoutpdf$.
We use duality to upper-bound $\mi(\matX;\matY)$ in~\eqref{eq:definition_capacity} as follows~\cite[Thm.~5.1]{lapidoth03-10a}:
\ifthenelse {\boolean{arxiv}}
{
\begin{IEEEeqnarray}{rCl}\label{eq:duality-expression}
\mi(\matX;\matY)&=& \Ex{\matX}{\relent{\condoutdist}{\outdist}}\notag\\
&\stackrel{(a)}{=}&\Ex{\matX}{\relent{\condoutdist}{\chosoutdist}} -\relent{\outdist}{\chosoutdist} \notag \\
&\stackrel{(b)}{\leq}& \Ex{\matX}{\relent{\condoutdist}{\chosoutdist}}\notag\\
&=& -\Ex{\outdist}{\log\lefto(\chosoutpdf(\matY)\right)}-\difent(\matY\given \matX).
\end{IEEEeqnarray}
}
{
\begin{IEEEeqnarray}{rCl}\label{eq:duality-expression}
\mi(\matX;\matY)&=& \Ex{\matX}{\relent{\condoutdist}{\outdist}}\notag\\
&\stackrel{(a)}{=}&\Ex{\matX}{\relent{\condoutdist}{\chosoutdist}} -\relent{\outdist}{\chosoutdist} \notag \\
&\stackrel{(b)}{\leq}& \Ex{\matX}{\relent{\condoutdist}{\chosoutdist}}\notag\\
&=& -\Ex{\outdist}{\log\chosoutpdf(\matY)}-\difent(\matY\given \matX).
\end{IEEEeqnarray}
}
Here,~(a) follows from Tops{\o}e's identity~\cite{topsoe67-a}, and (b) is a consequence of the nonnegativity of relative entropy~\cite[Thm.~2.6.3]{cover06-a}.
The conditional differential entropy $\difent(\matY \given \matX)$ in (\ref{eq:duality-expression}) is given by
\begin{IEEEeqnarray}{c}
\label{eq:h_y_x}
\difent(\matY \given \matX) = \rxant \sum\limits_{i=1}^{\txant} \Ex{}{\log \lefto(1+ \frac{\snr \|\vecx_i\|^2}{\txant}\right)} + \rxant \cohtime\log(\pi e).\IEEEeqnarraynumspace
\end{IEEEeqnarray}
To evaluate the first term on the RHS of~\eqref{eq:duality-expression}, we need to choose a specific output pdf $\chosoutpdf$.
%
%
Let us express $\matY$ in terms of its singular value decomposition (SVD)
\begin{equation}
\label{eq:SVD-Y}
\matY = \matU \matSigma \herm{ \matV}
\end{equation}
%
%
where $\matU \in \complexset^{\cohtime \times \minbound}$ and $\matV \in \complexset^{\rxant \times \minbound}$ ($\minbound$ is defined in \fref{tab:simple_functions}) are (truncated) unitary matrices, and $\matSigma=\diag\{\tp{[\sigma_1(\matY)\,\cdots\,\sigma_{\minbound}(\matY)]}\}$ contains the singular values of $\matY$ arranged in decreasing order.
To make the SVD unique, we shall assume that the diagonal entries of $\matU$ are real and non-negative.
Hence, \matV is an element of the complex \emph{Stiefel manifold}~$\stiefel(\rxant,\minbound)$~\cite{marques08-03a,zheng02-02a}, while \matU belongs to a \emph{submanifold}
 $\substiefel(\cohtime,\minbound)$ of $\stiefel(\cohtime,\minbound)$.
We put forward the following result about the volume of~$\stiefel(n,m)$ and~$\substiefel(n,m)$  for the case $n\geq m$ (see~\cite[Sec. V]{marques08-03a}) %
\begin{IEEEeqnarray*}{rCl}
	\abs{\stiefel(n,m)}&=&\frac{2^m \pi^{m n}}{\Gamma_{m}(n)};\,\,
	 \abs{\substiefel(n,m)}=\frac{\abs{\stiefel(n,m)}}{(2\pi)^{m}}=\frac{\pi^{m(n-1)}}{\Gamma_{m}(n)}.
\end{IEEEeqnarray*}
When~\inpdist is capacity-achieving, \fref{lem:structure_cap_ach_distr} and the Gaussianity of $\matH$ and $\matW$, imply that \matU and \matV are  uniformly distributed on $\substiefel(\cohtime,\minbound)$ and $\stiefel(\rxant,\minbound)$, respectively, and  independent of each other and of \matSigma.
We shall take an output pdf for which this property holds.
Furthermore, we take the first \txant singular values of \matY distributed as the ordered singular values of the noiseless channel output matrix $\sqrt{\snr/\txant}\cdot \matX\matH=\sqrt{\snr/\txant}\cdot\matPhi\matD\matH$, with $\matPhi$ unitary and isotropically distributed, and $\matD\distas \distoptD$.
By \fref{lem:BSTM_and_noiseless_output}, this implies that the first \txant singular values of \matY are distributed as the  singular values of
an $ \txant \times \minparam $ matrix with \iid $\jpg(0,\varsnr)$ entries, where
$\varsnr\define\rxant\cohtime\snr/(\txant\minparam)$.
We take the remaining $\minbound-\txant$ singular values distributed as the singular values of an independent $(\rxant-\txant)\times (\cohtime-\txant)$ matrix with \iid $\jpg(0,1)$ entries.
The intuition behind this choice is the following:
in the absence of the additive noise~\matW in~\eqref{eq:channel_IO}, the output matrix \matY has rank $\txant$; this suggests that, in the high-SNR regime, the smallest $\minbound-\txant$ singular values of \matY carry information about \matW only.
Summarizing, we take the pdf $\outpdf_{\sigma_1,\ldots,\sigma_{\minbound}}$ of the ordered singular values of \matY as follows\footnote{We shall indicate $\sigma_i(\matY)$ simply as $\sigma_i$ whenever no ambiguity occurs.}
\begin{IEEEeqnarray*}{rCl}
\outpdf_{\sigma_1,\ldots,\sigma_{\minbound}}(a_1,\cdots,a_{\minbound}) &=& \outpdf_{\sigma_1,\cdots,\sigma_\txant}(a_1,\cdots,a_\txant)\notag\\
&&\cdot\,\outpdf_{\sigma_{\txant+1},\cdots,\sigma_{\minbound}}(a_{\txant+1},\cdots,a_{\minbound})
\end{IEEEeqnarray*}
where
\begin{IEEEeqnarray}{rCl}
\IEEEeqnarraymulticol{3}{l}{
\outpdf_{\sigma_1,\cdots,\sigma_\txant}(a_1,\cdots,a_\txant)}\notag\\
 \,\,&=&\frac{k_1 e^{-\sum\nolimits_{i=1}^\txant a_i^2 / \varsnr}}{\varsnr^{\txant \minparam} }  \prod \limits_{i=1}^\txant a_i^{2(\minparam-\txant)+1}
\cdot\prod\limits_{i<j}^{\txant} \left( a_i^2-a_j^2\right)^2,\notag\\
&& 
 \hfill a_1>\dots > a_\txant\quad \label{eq:distribution-first-sv}
\end{IEEEeqnarray}
with $k_1\define 2^\txant \pi^{\txant(\txant-1)}/\bigl(\Gamma_\txant(\minparam)\Gamma_\txant(\txant)\bigr)$
and
%
\begin{IEEEeqnarray}{rCl}
\IEEEeqnarraymulticol{3}{l}{
\outpdf_{\sigma_{\txant+1},\cdots,\sigma_{\minbound}}(a_{\txant+1},\cdots,a_{\minbound})}\nonumber\\
 \quad &=&  k_2 e^{-\sum\nolimits_{i=\txant+1}^{\minbound} a_i^2} \!\!\!
 \prod \limits_{i=\txant+1}^{\minbound}\!\! a_i^{2(\maxbound-\minbound)+1}\cdot
  \!\!\biprod{\txant<i<j}{\minbound}\!\!\left( a_i^2-a_j^2\right)^2,\notag\\
  && 
  \hfill a_{\txant+1}> \dots> a_{\minbound} \quad \quad\label{eq:distribution-rest-sv}
\end{IEEEeqnarray}
%
%
%
with \maxbound defined in \fref{tab:simple_functions}, and
\begin{IEEEeqnarray*}{rCl}
k_2\define \frac{2^{\minbound-\txant}\pi^{(\minbound-\txant)(
\minbound-\txant-1)}}{\Gamma_{\minbound-\txant}(\maxbound-\txant)\Gamma_{\minbound-\txant}(\minbound-\txant)}.
\end{IEEEeqnarray*}
Here, both~\eqref{eq:distribution-first-sv} and~\eqref{eq:distribution-rest-sv} follow from~\cite[Thm.~2.17]{tulino04a} and the change of variable theorem.
We are now ready to evaluate the first term on the RHS of~\eqref{eq:duality-expression}.
Let
\begin{equation}\label{eq:jacobian}
J_{\maxbound,\minbound}(\sigma_1,\cdots,\sigma_{\minbound})
= \prod \limits_{i=1}^{\minbound} \sigma_i^{2(\maxbound-\minbound)+1} \cdot \biprod{i<j}{\minbound} \left( \sigma_i^2-\sigma_j^2\right)^2
\end{equation}
be the Jacobian of the SVD transformation~\cite[App. A]{zheng02-02a}.
The change of variables theorem yields
\begin{IEEEeqnarray}{rCl}
	-\Ex{}{\log\lefto( \chosoutpdf(\matY)\right)} &=& - \Ex{}{\log\lefto(\outpdf_{\matU,\matSigma,\matV}(\matU, \matSigma, \matV)\right)}
\notag\\
	&&+ \,\Ex{}{\log\lefto( J_{\maxbound,\minbound}(\sigma_1,\cdots,\sigma_{\minbound})\right) }\notag\\
	&=&  - \Ex{}{\log\lefto(\outpdf_{\matU}( \matU )\right)} -\Ex{}{\log\lefto(\outpdf_{\matV}(\matV)\right)}\notag\\
	&& -\,\Ex{}{\log\lefto(\outpdf_{\sigma_{\txant+1},\ldots,\sigma_{\minbound}}( \sigma_1,\ldots,\sigma_{\txant})\right)} \notag \\
	&& -\,\Ex{}{\log\lefto(\outpdf_{\sigma_{\txant+1},\ldots,\sigma_{\minbound}}(\sigma_{\txant+1},\ldots,\sigma_{\minbound})\right)}\notag\\
	&& +\, \Ex{}{\log\lefto( J_{\maxbound,\minbound}(\sigma_1,\cdots,\sigma_{\minbound})\right)}\label{eq:change-of-variable}		
\end{IEEEeqnarray}
where the second equality follows from the independence between \matU, \matV, and \matSigma.
Because \matU and \matV are uniformly distributed on the corresponding manifolds,
\begin{IEEEeqnarray}{rCl}
\label{eq:expectation_unitary_matrices1}
		  - \Ex{}{\log\lefto(\outpdf_{\matU}( \matU )\right)} &=& \log\abs{\substiefel(\cohtime,\minbound)}\\
		-\Ex{}{\log\lefto(\outpdf_{\matV}(\matV)\right)} &=&\log\abs{\stiefel(\rxant,\minbound)}.
\label{eq:expectation_unitary_matrices2}	
\end{IEEEeqnarray}
Substituting~\eqref{eq:distribution-first-sv},~\eqref{eq:distribution-rest-sv},~\eqref{eq:jacobian},~\eqref{eq:expectation_unitary_matrices1} and~\eqref{eq:expectation_unitary_matrices2} into~\eqref{eq:change-of-variable} we obtain after simple algebraic manipulations
\begin{IEEEeqnarray}{rCl}
	 -\Ex{}{\log\lefto( \chosoutpdf(\matY)\right)} &=&\txant\minparam \log (\varsnr) +
	 \log \lefto(\frac{\Gamma_{\txant}(\txant)\Gamma_{\txant}(\minparam)}{\Gamma_\txant(\rxant)\Gamma_\txant(\cohtime)}\right)\notag\\
	&& +\: \rxant\cohtime \log (\pi)   + k_3\Ex{}{\sum_{i=1}^{\txant}\log (\sigma_i^2)} \notag\\
	&& +\: \sum_{i=1}^{\txant}\sum_{j=\txant+1}^{\minbound} \Ex{}{\log\lefto((\sigma_i^2-\sigma_j^2)^2\right)}\notag\\
	&& +\:\frac{1}{\lambda}\sum\limits_{i=1}^{\txant}\Ex{}{\sigma_i^2}%
	+ \Ex{}{\sum\limits_{i=\txant+1}^{\minbound} \sigma_i^2}.\label{eq:second_step_computation_log_out}
\end{IEEEeqnarray}
Here, $k_3\define \maxbound-\minbound+\txant-\minparam$.
We next upper-bound the last three terms on the RHS of~\eqref{eq:second_step_computation_log_out}.
Using that the singular values are arranged in decreasing order we obtain
\begin{multline}\label{eq:the_third_last_term}
	  \sum_{i=1}^{\txant}\sum_{j=\txant+1}^{\minbound} \Ex{}{\log\lefto((\sigma_i^2-\sigma_j^2)^2\right)}\\
\leq 2(\minbound-\txant)\sum_{i=1}^{\txant}\Ex{}{\log (\sigma_i^2)}.
\end{multline}
For the second-last term, the power constraint~\eqref{eq:average-power-constraint} and the noise-variance normalization imply that
\begin{align}\label{eq:the_second_last_term}
   \frac{1}{\lambda}\sum\limits_{i=1}^{\txant}\Ex{}{\sigma_i^2} \leq \frac{\rxant\cohtime(\snr+1)}{\lambda}=\txant\minparam +\landauo(1), \quad \snr \to \infty
\end{align}
where we used that $\lambda=\rxant\cohtime\snr/(\txant\minparam)$.
Finally, to upper-bound the last term in~\eqref{eq:second_step_computation_log_out} we proceed as in~\cite[p.~377]{zheng02-02a} and obtain
\begin{align}\label{eq:the_last_term}
	 \Ex{}{\sum\limits_{i=\txant+1}^{\minbound} \sigma_i^2} \leq (\cohtime-\txant)(\rxant-\txant).
\end{align}
Substituting~\eqref{eq:the_third_last_term},~\eqref{eq:the_second_last_term}, and~\eqref{eq:the_last_term} into~\eqref{eq:second_step_computation_log_out}, and then~\eqref{eq:second_step_computation_log_out} and~\eqref{eq:h_y_x} into~\eqref{eq:duality-expression}, we get
\begin{IEEEeqnarray}{rCl}
	\IEEEeqnarraymulticol{3}{l}{\mi(\matX;\matY)}\notag\\
\:\:&\leq&
	\txant \minparam \log (\snr) -\txant\underbrace{(\rxant+\cohtime-\txant-\minparam)}_{=\maxparam}{}\notag\\
	&&+ \log \lefto(\frac{\Gamma_{\txant}(\txant)\Gamma_{\txant}(\minparam)}{\Gamma_\txant(\txant)\Gamma_\txant(\cohtime)}\right)	%
	+\txant \minparam \log \lefto(\frac{\rxant\cohtime}{\txant \minparam}\right) \notag\\
	 &&
	+\: (\cohtime-\txant-\minparam)\underbrace{\Ex{}{\sum\limits_{i=1}^{\txant}\log(\sigma_i^2)}}_{\define c_1(\snr)}{}\notag\\
	&&+\: \rxant\underbrace{\left(\Ex{}{\sum\limits_{i=1}^{\txant}\log(\sigma_i^2)} -\sum\limits_{i=1}^{\txant} \Ex{}{\log \lefto(1+ \frac{\snr \|\vecx_i\|^2}{\txant}\right)} \right)}_{\define c_2(\snr)}{}\notag\\
    && + \landauo(1),\quad \snr\rightarrow \infty.\label{eq:first_bound_on_mutual_inf}
\end{IEEEeqnarray}
To conclude the proof, we bound~$c_1(\snr)$ and $c_2(\snr)$ by exploiting that $\matX \notin \setK(\snr_0)$ \wpone.
Let $\matZ$ be a $(\cohtime-\txant) \times \rxant$ random matrix, independent of the channel matrix~\matH, and with \iid $\jpg(0,1)$ entries.
Given $\matX=\matPhi\matD$, the matrix $\herm{\matY}\matY$ has the same conditional distribution as~\cite[p.~377]{zheng02-02a}
\begin{align*}
	\underbrace{\herm{\matH}\left(\matI_{\txant} + \frac{\snr}{\txant}\matD^2\right)\matH}_{\define \matA}{} + \underbrace{\herm{\matZ}\matZ}_{\define \matB}{}.
\end{align*}
This property allows us to use Weyl's theorem~\cite[Thm.~4.3.1]{horn85a} to bound $c_1(\snr)$ as follows:
\begin{IEEEeqnarray}{rCl}\label{eq:first_bound_on_c1}
	c_1(\snr)&=&\Ex{\matX}{\Ex{\matY\given\matX}{\left.\sum_{i=1}^{\txant}\log\bigl(\eig{i}
	{\herm{\matY}\matY}\bigr) \right| \matX}}\notag\\
	&\leq& \Ex{\matX}{\Ex{\matH,\matZ}{\left.\sum_{i=1}^{\txant}\log\bigl(\eig{i}{\matA} +\eig{1}{\matB}\bigr) \right| \matX}} \notag\\
	&\leq&  \Ex{\matX}{\Ex{\matH}{\left.\sum_{i=1}^{\txant}\log\bigl(\eig{i}{\matA} +\underbrace{\Ex{\matZ}{\eig{1}{\matB}}}_{\define\eta}{}\bigr) \right| \matX}}.\IEEEeqnarraynumspace
\end{IEEEeqnarray}
Here, in the last step we used Jensen's inequality.
We next rewrite the argument in the expectation on the RHS of~\eqref{eq:first_bound_on_c1} in a more convenient form:
\begin{IEEEeqnarray}{rCl}
	\IEEEeqnarraymulticol{3}{l}{\sum_{i=1}^{\txant}\log\bigl(\eig{i}{\matA} +\eta\bigr)}\notag\\
    \,\, &\stackrel{(a)}{=}&  \logdet{\lefto(\matI_{\txant}+\frac{\snr}{\txant}\matD^2\right)\matH\herm{\matH}+\eta\matI_{\txant}} \notag\\
    &=& \log\det\lefto( \matI_{\txant}+\frac{\snr}{\txant}\matD^2\right)\notag  \\
	&& + \log\det\lefto(\matH\herm{\matH} +\diag\lefto\{\left[\eta\left(1+\frac{\snr}{\txant}\vecnorm{\vecx_1}^2\right)^{-1}\right.\vphantom {\left[ \eta\left(1+\frac{\snr}{\txant}\vecnorm{\vecx_{\txant}}^2\right)^{-1}\right]^{\mathsf{T}}} \right.\right.\notag\\
&&\quad\quad \quad\quad\quad\quad\quad \quad\quad \cdots\,\left.\left.\left.\eta\left(1+\frac{\snr}{\txant}\vecnorm{\vecx_{\txant}}^2\right)^{-1}\right]^{\mathsf{T}}\right\}\right) \notag\\
	&\stackrel{(b)}{\leq}& \log\det\lefto( \matI_{\txant}+\frac{\snr}{\txant}\matD^2\right) \notag\\
   &&+\underbrace{\log\det\lefto(\matH\herm{\matH} +\eta\left(1+\frac{\snr_0}{\txant }\right)^{-1}\matI_{\txant} \right)}_{\define \kappa(\matH,\snr_0)}.\label{eq:manipulation_of_argument}
\end{IEEEeqnarray}
Here,~(a) follows because $\herm{\matH}\left(\matI_{\txant}+(\snr/\txant)\matD^2\right)\matH$ and $\left(\matI_{\txant}+(\snr/\txant)\matD^2\right)\matH\herm{\matH}$ have the same $\txant$ nonzero eigenvalues~\cite[Thm.~1.3.20]{horn85a}, and (b) follows because $\matX \notin \setK(\snr_0)$ \wpone and
because for two matrices $\matA$ and $\matB$, if $\matA-\matB$ is positive semidefinite then $\det(\matA)\geq \det(\matB)$~\cite[Cor.~7.7.4]{horn85a}.
Substituting~\eqref{eq:manipulation_of_argument} into~\eqref{eq:first_bound_on_c1} we obtain
\begin{IEEEeqnarray}{rCl}
	c_1(\snr)&\leq& \Ex{}{\log\det\lefto(\matI_{\txant}+\frac{\snr}{\txant}\matD^2\right)}
	+  \Ex{}{\kappa(\matH,\snr_0)} \label{eq:useful_for_c2}\\
	&\leq& \txant \log\lefto(1+\frac{\cohtime\snr}{\txant}\right) +\Ex{}{\kappa(\matH,\snr_0)}\notag\\
	&=&\txant  \log\lefto(\frac{\cohtime\snr}{\txant}\right) +\Ex{}{\kappa(\matH,\snr_0)} +\landauo(1), \: \snr\to \infty.\IEEEeqnarraynumspace     \label{eq:bound_c1}
\end{IEEEeqnarray}
%
%
%
To bound $c_2(\snr)$ we use~\eqref{eq:useful_for_c2} and obtain
\begin{align}\label{eq:bound_c2}
	c_2(\snr)\leq  \Ex{}{\kappa(\matH,\snr_0)}.
\end{align}
Finally, substituting~\eqref{eq:bound_c1} and~\eqref{eq:bound_c2} into~\eqref{eq:first_bound_on_mutual_inf} we get
\begin{align}\label{eq:ub_capacity}
   \mi(\matX;\matY)\leq    \txant\left(\cohtime- \txant\right)\log (\snr) +\cohtime\cdot\constrho+ o(1),\: \snr\rightarrow \infty
\end{align}
where
\begin{IEEEeqnarray}{rCl}\label{eq:const_ub}
	\constrho &\define& \frac{1}{\cohtime}\log\lefto( \frac{\Gamma_{\txant}(\txant)\Gamma_{\txant}(\minparam)}{\Gamma_\txant(\rxant)\Gamma_\txant(\cohtime)}\right) + \txant\left(1-\frac{\txant}{\cohtime}\right)\log\lefto(\frac{\cohtime}{\txant}\right)\notag \\
	&& +\,\frac{\txant \minparam}{\cohtime} \log\lefto(\frac{\rxant}{\minparam}\right) + \frac{\maxparam}{\cohtime
	}\Bigl(\Ex{}{\kappa(\matH,\snr_0)}- \txant \Bigr).
\end{IEEEeqnarray}
Note that the RHS of~\eqref{eq:ub_capacity} does not depend on the choice of the input distribution.
Hence,~\eqref{eq:ub_capacity} is an upper bound on capacity as well.
Because $\matH$ has \iid Gaussian entries, and, hence, its singular values have finite differential entropy, we can apply~\cite[Lem.~6.7(b)]{lapidoth03-10a} combined with the dominated convergence theorem~\cite[p.~180]{grimmett01-a} and obtain
\begin{IEEEeqnarray*}{rCl}
	\lim_{\snr_0 \to \infty}\Ex{}{\kappa(\matH,\snr_0)} &=&\Ex{}{\lim_{\snr_0\to\infty}\kappa(\matH,\snr_0)}\\
&=&\Ex{}{\log\det(\matH\herm{\matH})}.
\end{IEEEeqnarray*}
Hence, \constrho in~\eqref{eq:const_ub}  can be made arbitrarily close to \const in~\eqref{eq:const_value_general}  by choosing $\snr_0$ sufficiently large.


\subsection{Lower Bound} 
\label{sec:lower_bound}
To obtain a capacity lower bound that matches the upper bound derived in \fref{sec:upper_bound}, we evaluate $I(\matX;\matY)$ for the BSTM input distribution introduced in \fref{sec:the_capacity_achieving_distribution_at_high_snr}.
More specifically, we proceed as follows.
Fix $\snr_0>0$ and let
\begin{IEEEeqnarray*}{rCl}
\diagsetK(\snr_0) \define \Big\{\matLambda = \diag&&\{\tp{[\lambda_1\,\cdots\,\lambda_\txant]}\} \sothat \\
&& \quad\quad0< \min\limits_{ m=1,\dots,\txant}\{ \lambda_m^2\} <\snr_0/\snr \Big\}.\
\end{IEEEeqnarray*}
Starting from \distoptD (see \fref{sec:the_capacity_achieving_distribution_at_high_snr}), we define the following family of probability distributions parameterized with respect to\footnote{Although \distoptDpar depends on both \snr and $\snr_0$, the choice of $\snr_0$ in the proof of the lower bound will turn out to be immaterial.} \snr
\begin{align}
\label{eq:input-norm-distribution}
\distoptDpar(\matLambda)= \begin{cases}
\frac{\distoptD(\matLambda)}{1-\Prob\lefto(\matD \in \diagsetK(\snr_0),\, \matD\distas\distoptD\right)}, &\text{if } \matLambda \notin \diagsetK(\snr_0) \\
 0, &\text{if } \matLambda \in \diagsetK(\snr_0).
\end{cases}
\end{align}
Note that $\distoptDpar(\matLambda)$ is supported outside $\diagsetK(\snr_0)$ and that $\lim_{\snr \to \infty} \distoptDpar(\matLambda)=\distoptD(\matLambda)$ for all~\matLambda.

\subsubsection{Preliminary Results} 
\label{sec:preliminary_results}

In \fref{lem:convergence_in_pdf} below, we establish that when $\matX=\matPhi\matD$ with $\matD\distas\distoptDpar$ and $\matPhi$ unitary and isotropically distributed, the joint pdf of the largest $\txant$ singular values of the output matrix \matY in~\eqref{eq:channel_IO} converges pointwise to the pdf of the nonzero singular values of $\sqrt{\snr/\txant}\cdot\matPhi\matD\matH$. Furthermore, the pdf of the remaining $\minbound-\txant$ singular values converge pointwise to the pdf of the singular values of an independent Gaussian matrix.
We remark that we implicitly used this property to construct the output distribution in \fref{sec:upper_bound}.
\begin{lem}\label{lem:convergence_in_pdf}
	Let $\matX=\matPhi\matD$ where \matPhi is unitary and isotropically distributed and $\matD\distas\distoptDpar$; let \matY as in~\eqref{eq:channel_IO}.
	Denote by $\sigma_1>\dots>\sigma_{\minbound}$ the singular values of \matY  and let
	\begin{IEEEeqnarray}{c}\label{eq:normalized_singular_values}
		\altvecsigma=\tp{\left[\bigl(\sqrt{\txant/\snr}\bigr) \sigma_1\, \cdots\, \bigl(\sqrt{\txant/\snr}\bigr)\sigma_\txant \,\,\,\sigma_{\txant+1}\, \cdots\, \sigma_{\minbound}\right]}.\IEEEeqnarraynumspace
	\end{IEEEeqnarray}
	The pdf of \altvecsigma converges pointwise as $\snr\to \infty$ to the pdf of a vector $\vecu\in \complexset^{\minbound}$ whose first \txant entries are distributed as the  ordered nonzero singular values of $\matD\matH$, with $\matD\distas\distoptD$ and $\matH$ as in~\eqref{eq:channel_IO}, and whose remaining $\minbound-\txant$ entries are distributed  as the nonzero  singular values of an independent $(\cohtime - \txant) \times (\rxant-\txant)$ random matrix with i.i.d. $\jpg(0,1)$ entries.
\end{lem}
\begin{IEEEproof}
	\ifthenelse {\boolean{arxiv}}
	{
	See \fref{app:proof_of_lemma_lem:convergence_in_pdf}.
	}
	{
	The rather technical proof is omitted and can found in~\cite[App.~A]{yang12-06a}.
	}	
\end{IEEEproof}
Note that by Scheff\'{e}'s Theorem~\cite{scheffe47}, pointwise convergence of pdfs implies convergence in distribution of \altvecsigma to \vecu. This weaker convergence result (which is not sufficient to establish our capacity lower bound) has been already pointed out (without proof) in~\cite[Lem.~16]{zheng02-02a}.

%
%
In \fref{lem:useful_asyptotic_results} below we collect four asymptotic results regarding the differential entropy and the expected logarithm of the entries of \altvecsigma in~\eqref{eq:normalized_singular_values}  that we shall need in the proof of the lower bound.
\begin{lem}\label{lem:useful_asyptotic_results}
	Let $\altvecsigma=\tp{[\altsigma_1\, \cdots\, \altsigma_{\minbound}]}$  and $\vecu=\tp{[u_1\,\cdots\,u_{\minbound}]}$ as in \fref{lem:convergence_in_pdf}.
	Then
	\begin{enumerate}
	\item 
	$\difent(\altvecsigma) = \difent(\vecu) + \landauo(1),~\snr\rightarrow\infty$
	\item 
	$\Ex{}{\log (\altsigma_i)} = \Ex{}{\log( u_i)}+ \landauo(1),~\snr\rightarrow\infty$, $1\leq i\leq \minbound$
	\item 
	$\Ex{}{\log\lefto( \altsigma_i^2-\altsigma_j^2\right)} = \Ex{}{\log (u_i^2-u_j^2)}+ \landauo(1),~\snr\rightarrow\infty$, $1\leq i<j\leq \minbound$
	\item
	$\Ex{}{\log\lefto( \altsigma_i^2-\txant \altsigma_j^2/\snr\right)} = \Ex{}{\log (u_i^2)}+ \landauo(1),~\snr\rightarrow\infty$, $1\leq i<j\leq \minbound$.
	\end{enumerate}
\end{lem}
\begin{IEEEproof}
	\ifthenelse {\boolean{arxiv}}
	{
	See \fref{app:proof_of_lemma_lem:useful_asyptotic_results}.
	}
	{
	The proof, which is again technical, is omitted and can be found in~\cite[App.~B]{yang12-06a}.
	}
\end{IEEEproof}

\subsubsection{The Actual Bound} 
\label{sec:the_actual_bound}
We evaluate the mutual information
\begin{align}\label{eq:mi_MIMO}
	I(\matX;\matY)=\difent(\matY)-\difent(\matY\given\matX)
\end{align}
in~\eqref{eq:definition_capacity} for $\matX=\matPhi\matD$ with \matPhi unitary and isotropically distributed and $\matD\distas\distoptDpar$.
The second term on the RHS of~\eqref{eq:mi_MIMO} is given by
\begin{IEEEeqnarray}{rCl}
\label{eq:lb-difent-Y-X}
\difent(\matY\given\matX) &=& \rxant \cohtime\log(\pi e)+\rxant \sum\limits_{i=1}^{\txant} \Ex{}{\log \lefto(1+ {\snr \|\vecx_i\|^2}/{\txant}\right)}\notag\\
& =& \rxant \cohtime\log(\pi e) + \rxant \sum\limits_{i=1}^{\txant} \Ex{}{\log (\snr d_i^2 /\txant)} \notag\\
&&+\: \rxant \sum\limits_{i=1}^{\txant} \Ex{}{\log\lefto(1+\txant/{(\snr d_i^2)}\right)}\notag\\
& \stackrel{(a)}{=}& \rxant \cohtime\log(\pi e) + \txant\rxant \log\lefto({\snr}/{\txant}\right) \notag\\ &&+\: \rxant \Ex{}{\logdet{ \matD^2}} + o(1),\quad \snr\rightarrow\infty.
\end{IEEEeqnarray}
Here, (a) follows because $\snr d^2_i\geq \snr_0$ \wpone, and hence, $0\leq \log\bigl(1+\txant /(\snr d_i^2)\bigr)\leq \log(1+\txant/\snr_0)$ \wpone, which implies that
\begin{IEEEeqnarray*}{rCl}
\lim\limits_{\snr\rightarrow\infty}\Ex{}{\log\lefto(1+ \frac{\txant}{\snr d_i^2}\right)} &=& \Ex{}{\lim\limits_{\snr\rightarrow\infty} \log\lefto(1+\frac{\txant}{\snr d_i^2}\right)} =0
\end{IEEEeqnarray*}
as a consequence of the dominated convergence theorem~\cite[p.~180]{grimmett01-a}.
We shall compute $\difent(\matY)$ in SVD coordinates [cf.,~\eqref{eq:SVD-Y}] as follows:
\begin{IEEEeqnarray}{rCl}\label{eq:lb-difent-Y}
\difent(\matY) &\stackrel{(a)}{=}& \underbrace{\difent(\matU)}_{=\log |\substiefel(\cohtime,\minbound)|}{}+\underbrace{\difent(\matV)}_{=\log|\stiefel(\rxant,\minbound)|}{}+\difent(\sigma_1,\ldots,\sigma_{\minbound}) \notag\\
&& +\: \Ex{}{\log\lefto( J_{\maxbound,\minbound}(\sigma_1,\ldots,\sigma_{\minbound})\right)}\notag\\
&\stackrel{(b)}{=}&\log |\substiefel(\cohtime,\minbound)| + \log|\stiefel(\rxant,\minbound)| + \frac{\txant}{2}\log\lefto(\frac{\snr}{\txant}\right)\notag\\
&& + \:\difent(\altvecsigma) + \Ex{}{\log \lefto(J_{\maxbound,\minbound}(\sigma_1,\ldots,\sigma_{\minbound})\right)}.
\end{IEEEeqnarray}
Here,~(a) follows because the isotropic distribution of \matPhi and the Gaussianity of \matH and \matW  imply that \matU and \matV are uniformly distributed on $\substiefel(\cohtime,\minbound)$ and $\stiefel(\rxant,\minbound)$, respectively, and independent of \matSigma;
%
In (b), we used~\eqref{eq:normalized_singular_values} and that $\difent(\matA\vecx) = \difent(\vecx)+\log\det(\matA)$ for a random vector $\vecx$ and a deterministic matrix \matA \cite[Eq.~(8.71)]{cover06-a}.
It is convenient to express also the Jacobian $J_{\maxbound,\minbound}$ in~\eqref{eq:lb-difent-Y} in terms of \altvecsigma.
Using~\eqref{eq:jacobian} and~\eqref{eq:normalized_singular_values} we obtain
\begin{IEEEeqnarray}{rCl}
	\IEEEeqnarraymulticol{3}{l}{
	\Ex{}{\log \lefto(J_{\maxbound,\minbound}(\sigma_1,\ldots,\sigma_{\minbound})\right)}}\notag\\
\:\: &=& k_4\log\lefto(\frac{\snr}{\txant}\right) + \sum\limits_{i=1}^{\txant}\Ex{}{\log \lefto(\tilde{\sigma}_i^{2(\maxbound-\minbound)+1}\right)} \notag\\
&& + \sum\limits_{i<j}^{\txant}\Ex{}{\log \lefto(\left(\tilde{\sigma}_i^2-\tilde{\sigma}_j^2\right)^2\right)}\notag\\
	&& + \sum\limits_{i=\txant+1}^{\minbound} \Ex{}{\log \lefto(\tilde{\sigma}_i^{2(\maxbound-\minbound)+1}\right)} \notag\\
&&+ \sum\limits_{\txant<i<j}^{\minbound}\Ex{}{\log \lefto(\left(\tilde{\sigma}_i^2-\tilde{\sigma}_j^2\right)^2\right)}\notag\\
	   && + \sum\limits_{i=1}^{\txant}\sum\limits_{j=\txant+1}^{\minbound} \Ex{}{\log \lefto(\left(\tilde{\sigma_i}^2-{\txant \tilde{\sigma}_j^2}/{\snr}\right)^2\right)}\label{eq:lb-jacobian}
\end{IEEEeqnarray}
where $k_4\define\txant(\maxbound+\minbound-\txant-1/2)$.
Substituting~\eqref{eq:lb-jacobian} into (\ref{eq:lb-difent-Y}), and using Lemma~\ref{lem:useful_asyptotic_results}, we obtain
\begin{IEEEeqnarray}{rCl}
\label{eq:lb-difent-Y-2}
\difent(\matY) &=& \log |\substiefel(\cohtime,\minbound)| + \log |\stiefel(\rxant,\minbound)|\notag\\
&& +\: \txant(\maxbound+\minbound-\txant)\log\lefto({\snr}/{\txant}\right)
 + h(u_1,\ldots,u_\txant) \notag\\
 && + \sum\limits_{i=1}^{\txant}\Ex{}{\log \lefto(u_i^{2(\maxbound+\minbound-2\txant)+1}\right)} \notag\\
 && + \sum\limits_{i<j}^{\txant}\Ex{}{\log\lefto(\left(u_i^2-u_j^2\right)^2\right)} + h(u_{\txant+1},\ldots,u_{\minbound})\notag\\
 && + \sum\limits_{i=\txant+1}^{\minbound} \Ex{}{\log\lefto( u_i^{2(\maxbound-\minbound)+1}\right)}\notag\\
 && + \sum\limits_{\txant<i<j}^{\minbound}\Ex{}{\log \lefto(\left(u_i^2-u_j^2\right)^2\right)} + o(1),\,\snr\rightarrow\infty.\IEEEeqnarraynumspace
\end{IEEEeqnarray}
We next evaluate the terms on the RHS of~\eqref{eq:lb-difent-Y-2} by proceeding as follows.
By Lemmas~\ref{lem:convergence_in_pdf} and~\ref{lem:BSTM_and_noiseless_output}, $\{u_1,\dots,u_\txant\}$ are jointly distributed as the singular values of an $\txant \times \minparam$ Gaussian random matrix \matG with \iid $\jpg(0,\sqrt{\rxant\cohtime/\minparam})$ entries.
Evaluating $\difent(\matG)$ in the SVD coordinate system, we get
\begin{IEEEeqnarray}{rCl}
\label{eq:lb-difent-G}
\difent(\matG)
& =& \log|\widetilde{\setS}(\txant,\txant)|+\log|\setS(\minparam,\txant)| + \difent(u_1,\ldots,u_\txant)
\notag\\
&& + \sum\limits_{i=1}^{\txant} \Ex{}{\log \lefto(u_i^{2(\minparam-\txant)+1}\right)} \notag\\
&& + \sum\limits_{i<j}^{\txant}\Ex{}{\log \lefto(\left(u_i^2-u_j^2\right)^2\right)}.
\end{IEEEeqnarray}
Similarly, by \fref{lem:convergence_in_pdf}, $\{u_{\txant+1},\dots, u_{\minparam} \}$ are jointly distributed as the singular values of a $(\cohtime-\txant)\times(\rxant-\txant)$ random Gaussian matrix \altnoisemat with \iid $\jpg(0,1)$ entries.
Thus,
\begin{IEEEeqnarray}{rCl}
\label{eq:lb-difent-W}
\difent(\altnoisemat)
&=& \log\lefto|\substiefel(\cohtime-\txant, \minbound-\txant)\right| + \log \lefto|\stiefel(\rxant-\txant,\minbound- \txant)\right| \notag\\
&& +\:\difent(u_{\txant+1},\ldots,u_{\minbound})  + \sum\limits_{i=\txant+1}^{\minbound} \Ex{}{\log \lefto(u_i^{2(\maxbound-\minbound)+1}\right)}  \notag\\
&&+ \sum\limits_{\txant<i<j}^{\minbound}\Ex{}{\log \lefto(\left(u_i^2-u_j^2\right)^2\right)}.
\end{IEEEeqnarray}
Substituting \fref{eq:lb-difent-G} and \fref{eq:lb-difent-W} into \fref{eq:lb-difent-Y-2}, and then \fref{eq:lb-difent-Y-X} and \fref{eq:lb-difent-Y-2} into (\ref{eq:mi_MIMO}), we obtain
\begin{IEEEeqnarray}{rCl}
\IEEEeqnarraymulticol{3}{l}{
\mi(\matX;\matY)} \notag\\
\,\,&=& \txant(\cohtime-\txant)\log\lefto(\frac{\snr}{\txant}\right) + \difent(\matG) + \difent(\altnoisemat)\notag\\
&& + \underbrace{\sum\limits_{i=1}^{\txant}\Ex{}{\log \lefto(u_i^{2(\maxbound+\minbound-\txant-\minparam)}\right)} - \rxant\Ex{}{\logdet{ \matD^2}} }_{\define \alpha}{} \notag\\
&&+ \log (k_5) - \rxant\cohtime \log(\pi e)+o(1),\quad\snr\rightarrow\infty\label{eq:lb-mi}
\end{IEEEeqnarray}
where $k_5\define|\substiefel(\cohtime,\minbound)|\cdot|\stiefel(\rxant,\minbound)|/\Bigl[|\substiefel(\cohtime-\txant, \minbound-\txant)| \cdot  |\stiefel(\rxant-\txant,\minbound- \txant)| \cdot |\substiefel(\txant,\txant)| \cdot |\stiefel(\minparam,\txant)|\Bigr]$.
The term denoted by $\alpha$ in~\eqref{eq:lb-mi} can be simplified as follows:
\begin{IEEEeqnarray}{rCl}
\label{eq:lb-term-alpha}
\alpha &\stackrel{(a)}{=}& \maxparam\Ex{}{\sum\limits_{i=1}^{\txant}\log\lefto( u_i^2\right)} - \rxant\Ex{}{\logdet{\matD^2}}\notag\\
&\stackrel{(b)}{=}&\maxparam \Ex{}{\logdet{\matD^2\matH\herm{\matH}}} -\rxant\Ex{}{\logdet{\matD^2}}\notag\\
&=& (\maxparam-\rxant) \Ex{}{\logdet{\matD^2}} + \maxparam \Ex{}{\logdet{\matH\herm{\matH}}}\notag\\
& \stackrel{(c)}{=}& (\maxparam-\rxant)\txant \log (\cohtime) + \maxparam \Ex{}{\logdet{\matH\herm{\matH}}}.
\end{IEEEeqnarray}
Here, in (a) we used that $\maxparam=\maxbound+\minbound-\txant-\minparam$,~(b) follows from \fref{lem:convergence_in_pdf}, and~(c) holds because when $\cohtime\geq \txant +\rxant$ we have that $\matD=\sqrt{\cohtime}\cdot\matI_{\txant}$, and when $\cohtime< \txant +\rxant$ we have that $\maxparam-\rxant=0$.
Finally, substituting \fref{eq:lb-term-alpha} into \fref{eq:lb-mi},  we get after straightforward algebraic manipulations
\begin{align*}
\mi(\matX;\matY) & =\txant(\cohtime-\txant)\log(\snr) +\cohtime \cdot \const  +\landauo(1),~\,\,\snr\rightarrow\infty
\end{align*}
where $\const$ is given in~\eqref{eq:const_value_general}. This concludes the proof.




\section{Conclusions} 
\label{sec:conclusions}
It was shown in~\cite{zheng02-02a} that USTM achieves the high-SNR capacity of a Rayleigh block-fading MIMO channel in the regime where the channel's coherence time \cohtime is larger or equal to the sum of the number of transmit antennas \txant and receive antennas \rxant.
In the same paper, it was also conjectured that when $\cohtime<\txant+\rxant$, a situation relevant for large-MIMO systems, USTM is no longer optimal.
In this paper, we prove this conjecture.
Specifically, we establish that USTM is not capacity-achieving when $\cohtime<\txant+\rxant$  by determining the input distribution (which we refer to as BSTM) that achieves capacity at high SNR.
The corresponding capacity-achieving input signal is the product of a unitary isotropically distributed matrix and a diagonal matrix whose nonzero entries are distributed as the square-root of the eigenvalues of a Beta-distributed matrix of appropriate size.
The analytical and numerical results reported in \fref{sec:gain_of_BSTM} illustrate that the rate gain determined by using BSTM instead USTM grows logarithmically in the number of receive antennas $\rxant$, and can be as large as $13\%$ for practically relevant SNR values, when $\rxant\gg \cohtime$ and $\txant=\floor{\cohtime/2}$.

\appendices

\section{Proof of Lemma~\ref{lem:convergence_in_pdf}} 
\label{app:proof_of_lemma_lem:convergence_in_pdf}
Throughout this appendix, we shall focus for simplicity on the case $\cohtime\leq \rxant$.
We shall, however, outline the additional steps needed to generalize the proof to the case $\cohtime>\rxant$.
Let $\pdfoptDpar$ and $\pdfoptD$ be the pdfs corresponding to the probability distributions \distoptDpar and \distoptD, respectively (such pdfs exist when $\cohtime\leq \rxant$).
Let  $\genericpdf_{\matY\given \matD}$ denote the conditional pdf of \matY  given \matD.
Denote by \pdfaltsigma and \pdfu the pdf of \altvecsigma and \vecu, respectively.
Finally, denote by \pdfaltsigmacond and \pdfucond  the conditional pdf of \altvecsigma and \vecu given \matD.
The proof consists of the following three steps:
\begin{enumerate}
	\item We first obtain a closed-form expression for $\genericpdf_{\matY\given \matD}$, thus generalizing the result obtained in~\cite[Sec.~III.A]{hassibi02-06a} (for the special case of $\matD$ being a scaled identity matrix) to arbitrary diagonal matrices.
	This result is of independent interest.
	\item We then calculate \pdfaltsigmacond from  $\genericpdf_{\matY\given \matD}$ and show that $\pdfaltsigmacond$ converges pointwise to \pdfucond as $\snr\to\infty$.
	\item Finally, we show that
	\begin{align}\label{eq:bounded_by_const_app1}
		\abs{\pdfaltsigmacond(\altvecsigmaarg\given\matDarg)\cdot  \pdfoptDpar(\matDarg)}\leq \constapp
	\end{align}
	 where \constapp is a finite constant that does not depend on $\altvecsigmaarg$ and $\matDarg$, i.e., the bound is uniform in both $\altvecsigmaarg$ and $\matDarg$.
	As $\matD\distas\distoptDpar$ implies that $\matD$ has compact support,
	we can invoke the dominated convergence theorem~\cite[Thm.~1.34]{rudin87a} and conclude that
	\begin{IEEEeqnarray*}{rCl}
	\lim\limits_{\snr\rightarrow\infty} \pdfaltsigma(\altvecsigmaarg)&=&
	\lim\limits_{\snr\rightarrow\infty} \int \pdfaltsigmacond(\altvecsigmaarg\given \matDarg) \pdfoptDpar(\matDarg) d\matDarg\notag\\
	&=&\int \lim_{\snr \to \infty}\left[\pdfaltsigmacond(\altvecsigmaarg\given \matDarg) \pdfoptDpar(\matDarg)\right] d\matDarg\notag\\
	&=&\int \pdfucond(\altvecsigmaarg\given \matDarg)\pdfoptD(\matDarg) d\matDarg= \pdfu(\altvecsigmaarg).
	\end{IEEEeqnarray*}
\end{enumerate}
\subsection{Step 1} 
\label{par:step_1}
Set $\altsnr\define\snr/\txant$. Since \matY is conditionally Gaussian given \matX,
\begin{align*}
	 \genericpdf_{\matY\given\matX}(\matY\given\matX) = \frac{1}{\pi^{\rxant\cohtime}}\cdot\frac{\exp\lefto[-\tr\lefto(\herm{\matY}\left(\altsnr\matX\herm{\matX} + \matI_\cohtime\right)^{-1}\matY\right)\right]}{\det\lefto(\altsnr\matX\herm{\matX} + \matI_\cohtime\right)^\rxant}.
\end{align*}
To obtain $\genericpdf_{\matY\given \matD}$ from  $\genericpdf_{\matY\given\matX}$, it is convenient to consider the eigenvalue decomposition of $\matY\herm{\matY}$:
\begin{equation}\label{eq:eig-dec-appA}
\matY\herm{\matY} = \widetilde{\matU}\underbrace{\left(
                                 \begin{array}{c c}
                                   \matSigma^2 &  \mathbf{0}_{\minbound \times(\cohtime-\minbound)} \\
                                    \mathbf{0}_{(\cohtime-\minbound)\times \minbound} & \mathbf{0}_{\cohtime-\minbound} \\
                                 \end{array}
                               \right)}_{\define \altmatSigma}
\herm{\widetilde{\matU}}.
\end{equation}
Here, $\widetilde{\matU}$ is a $\cohtime\times\cohtime$ unitary matrix, and \matSigma, defined in~\eqref{eq:SVD-Y}, contains the singular values $\sigma_1>\dots>\sigma_{\minbound}$ of \matY.
Set now $\matLambda\define(\altsnr^{-1}\matD^{-2}+\matI_{\txant})^{-1}$ and recall that $\matX=\matPhi\matD$, where $\matPhi$ is unitary and isotropically distributed, and, hence, uniformly distributed on $\stiefel(\cohtime,\txant)$.
Proceeding as in~\cite[Sec.~III]{hassibi02-06a},
\begin{IEEEeqnarray}{rCl}
\label{eq:cond_pdf_with_tricky_integral}
	 \genericpdf_{\matY\given\matD}(\matY\given\matD) &=& \frac{1}{\abs{\stiefel(\cohtime,\txant)}} \int \genericpdf_{\matY\given \matX}(\matY\given\matPhi\matD)d\matPhi \notag\\
	&=&   \frac{1}{\pi^{\rxant\cohtime}}\cdot
	\frac{\exp\lefto[-\tr\lefto(\herm{\matY}\matY\right)\right]}{\det\lefto(\altsnr\matD^2 + \matI_\txant\right)^\rxant} \cdot \frac{1}{\abs{\stiefel(\cohtime,\txant)}}\notag\\
 && \cdot\,\int\nolimits_{\stiefel(\cohtime,\txant)} \exp\lefto[\tr\lefto(\altmatSigma\matPhi\matLambda\herm{\matPhi}\right)\right]d\matPhi .
\end{IEEEeqnarray}
The integral on the RHS of~\eqref{eq:cond_pdf_with_tricky_integral} is computed in closed-form in~\cite[Sec. III.A]{hassibi02-06a} for the special case $\matD=\sqrt{\cohtime}\cdot\matI_{\txant}$, which corresponds to USTM.
We shall next evaluate this integral (and, hence, $\genericpdf_{\matY\given\matD}$) in closed-form for an arbitrary diagonal matrix \matD.
We start by observing that the integral under examination resembles the well-known Itzykson-Zuber integral~\cite[Eq.~(3.2)]{itzykson80-a}, with the crucial differences that, in our case, the integration is performed over the Stiefel manifold $\stiefel(\cohtime,\txant)$ instead of the \emph{unitary group} $\unitgroup(\cohtime)\define \stiefel(\cohtime,\cohtime)$.
Let $\altmatPhi=[\matPhi\,\, \matPhi_{\bot}]$ where $\matPhi_{\bot}$ is a $\cohtime \times (\cohtime-\txant)$ matrix chosen so that $\altmatPhi$ is unitary, i.e., $\altmatPhi \in \unitgroup(\cohtime)$.
Then~\cite[Eq.~(5)]{onatski08-04a}
\begin{IEEEeqnarray}{rCl}
\label{eq:new_integral_app1}
	\int_{\stiefel(\cohtime,\txant)}\!\!\!\!\! e^{\tr\lefto(\altmatSigma\matPhi\matLambda\herm{\matPhi}\right)} d\matPhi  = \frac{1}{\abs{\unitgroup(\cohtime -\txant)}}
	\int_{\unitgroup(\cohtime)}\!\!\!e^{\tr\lefto(\altmatSigma\matPhi\matLambda\herm{\matPhi}\right)} d\altmatPhi.\IEEEeqnarraynumspace
\end{IEEEeqnarray}
The assumption $\cohtime\leq\rxant$ entails that the nonzero entries of the diagonal matrix \matD are distinct (see~\fref{sec:the_capacity_achieving_distribution_at_high_snr}); hence, the nonzero entries of the diagonal matrix \matLambda are distinct as well.
%
Furthermore, when $\cohtime \leq \rxant$ we have that $\minbound=\cohtime$ (and $\maxbound=\rxant$) and, hence, $\matDelta=\matSigma^2$ [see~\eqref{eq:eig-dec-appA}].
Starting from $\matLambda=\diag\{\tp{[\lambda_1\,\cdots\,\lambda_{\txant}]}\}$, we next define the following full-rank $\cohtime\times\cohtime$ diagonal matrix:
\begin{align*}
 \matLambda_{\epsilon}\define\diag\{\tp{[\lambda_1\,\cdots\,\lambda_{\txant}\,\,\epsilon'_{\txant+1}\,\cdots\,\epsilon'_{\cohtime}]}\}.
\end{align*}
Here, $\epsilon'_{\txant+1},\dots,\epsilon'_{\cohtime}$ are nonnegative real numbers chosen so that the nonzero entries of $\matLambda_{\epsilon}$ are distinct.
As the unitary group $\unitgroup(\cohtime)$ is compact,
\begin{align}\label{eq:reduction_to_itzykson_zuber}
	\int_{\unitgroup(\cohtime)}\!\!e^{\tr\lefto(\matSigma^2\matPhi\matLambda\herm{\matPhi}\right)} d\altmatPhi
	=
	\lim_{\epsilon'_{\txant+1}\to 0,\dots,\epsilon'_{\cohtime}\to 0} \int_{\unitgroup(\cohtime)}\!\!e^{\tr\lefto(\matSigma^2\altmatPhi\matLambda_{\epsilon}\herm{\altmatPhi}\right)} d\altmatPhi.
\end{align}
The argument of the $\lim$ operator on the RHS of~\eqref{eq:reduction_to_itzykson_zuber} is the Itzykson-Zuber integral.
Hence, by~\cite[Eq.~(3.4)]{itzykson80-a} we get\footnote{Note that---differently from our setup---in~\cite[Eq.~(3.4)]{itzykson80-a} the Haar measure on the unitary group is normalized.}
\begin{IEEEeqnarray}{rCl}
{\int\limits_{\unitgroup(\cohtime)}e^{\tr\lefto(\matSigma^2\altmatPhi\matLambda_{\epsilon}\herm{\altmatPhi}\right)} d\altmatPhi} &=& \frac{\abs{\unitgroup(\cohtime)} \cdot \prod\limits_{i=1}^{T} \Gamma(i) \cdot \det(\matA)}{\prod\limits_{i<j}^{\cohtime}(\sigma^2_i-\sigma^2_j)\cdot\prod\limits_{i<j}^{\cohtime}(\altlambda_i-\altlambda_j)}.\IEEEeqnarraynumspace \label{eq:itzykson_zuber}
\end{IEEEeqnarray}
Here, $\{\altlambda_j\}_{j=1}^{\cohtime}$ are the diagonal entries of $\matLambda_{\epsilon}$, and \matA is a $\cohtime \times \cohtime$ matrix defined as follows: $[\matA]_{i,j}=\exp(\sigma^2_j\altlambda_i)$.
We next compute the limit  $\epsilon'_{\txant+1}\to 0,\dots,\epsilon'_{\cohtime}\to 0$ of the RHS of~\eqref{eq:itzykson_zuber} using l'H\^{o}pital's Theorem, substitute the final result into~\eqref{eq:new_integral_app1}, and obtain~\cite[Lem.~5]{ghaderipoor12}
\begin{IEEEeqnarray}{rCl}
\IEEEeqnarraymulticol{3}{l}
   {\int\nolimits_{\stiefel(\cohtime,\txant)}e^{\tr\lefto(\altmatSigma\matPhi\matLambda\herm{\matPhi}\right)} d\matPhi}\notag\\
   \:\: &=&	
	\underbrace{\frac{\abs{\unitgroup(\cohtime)}}{\abs{\unitgroup(\cohtime-\txant)}}}_{=\abs{\stiefel(\cohtime,\txant)}}{}
	\cdot  \frac{ \det (\matM) \det\lefto(\matLambda^{ \txant - \cohtime}\right)\cdot\prod\limits_{i=\cohtime-\txant+1}^{\cohtime} \!\!\!\Gamma(i)}{\prod\limits_{i<j}^{\cohtime}(\sigma_i^2 - \sigma_j^2)\cdot\prod\limits_{i<j}^{\txant}(\lambda_i-\lambda_j)}\IEEEeqnarraynumspace\label{eq:integral_solved}
\end{IEEEeqnarray}
with \matM being a $\cohtime \times \cohtime$ matrix defined as follows:
\begin{align*}
[\matM]_{i,j} = \begin{cases}
e^{\lambda_i\sigma_j^2}, & 1\leq i\leq \txant,\,\, 1 \leq j\leq \cohtime\\
%
%
%
%
\sigma_j^{2(\cohtime-i)},
& \txant<i \leq \cohtime,\,\, 1\leq j\leq \cohtime.
%
%
\end{cases}
\end{align*}
Substituting~\eqref{eq:integral_solved} into~\eqref{eq:cond_pdf_with_tricky_integral} and using that $\matLambda=\left(\altsnr^{-1}\matD^{-2} + \matI_\txant\right)^{-1}$, we obtain the following closed-form expression for the conditional pdf $\genericpdf_{\matY\given\matD}$:
\begin{IEEEeqnarray}{rCl}
\IEEEeqnarraymulticol{3}{l}{\genericpdf_{\matY\given\matD}(\matY\given\matD)}\notag\\
 \quad&=&
\frac{1}{\pi^{\rxant\cohtime}}\cdot \prod\limits_{i=\cohtime-\txant+1}^{\cohtime} \!\!\!\Gamma(i)
\cdot\frac{\exp\lefto[-\tr\lefto(\herm{\matY}\matY\right)\right]}{\det\lefto(\altsnr\matD^2 + \matI_\txant
\right)^{\rxant}}\notag\\
&&\cdot\,\frac{\det(\matM) \det\lefto(\matLambda^{\txant - \cohtime}\right)}{\prod\limits_{i<j}^{\cohtime}(\sigma_i^2 - \sigma_j^2)\cdot\prod\limits_{i<j}^{\txant}(\lambda_i-\lambda_j)}\notag\\
&=&\frac{\altsnr^{-\txant
(\txant-1)/2}}{\pi^{\rxant\cohtime}}\cdot \!\!\!\prod\limits_{i=\cohtime-\txant+1}^{\cohtime}\!\!\! \Gamma(i)
\cdot\frac{ \det \lefto(\altsnr^{-1}\matD^{-2} + \matI_\txant
\right)^{\cohtime-\txant} }{\det\lefto(\altsnr\matD^2 + \matI_\txant\right)^{\rxant-\txant
+1}}\notag\\
&&\cdot\,\frac{\exp\lefto(-\sum\nolimits_{i=1}^{\minbound}\sigma_i^2\right)}{\prod\limits_{i<j}^{\cohtime}(\sigma_i^2 - \sigma_j^2)}\cdot\frac{\det(\matM)}{\prod\limits_{i<j}^{\txant
}(d_i^2-d_j^2)}.\IEEEeqnarraynumspace\label{eq:conditional-pdf-Y-closed-form}
\end{IEEEeqnarray}
We remark that~\eqref{eq:conditional-pdf-Y-closed-form} holds under the assumption that $\cohtime\leq\rxant$, which ensures that the $\{d_i\}_{i=1}^{\txant}$ are all distinct.

When $\cohtime>\rxant$, we have that $d_1=\dots=d_l=\sqrt{\cohtime\rxant/\minparam}$, where $l=\cohtime - \maxparam = \min\{\txant,\cohtime-\rxant\}$  (see~\fref{sec:the_capacity_achieving_distribution_at_high_snr}).
Hence, $\lambda_1=\cdots=\lambda_l=\lambda\define [\minparam/(\cohtime\rxant\altsnr)+1]^{-1}$.
Let   in this case
\begin{align*}
	\matLambda'_{\epsilon}\define \diag\{\tp{[\lambda+\epsilon'_1\,\cdots\,\lambda+\epsilon'_l\,\,\, \lambda_{l+1}\,\cdots\, \lambda_{\txant}\,\,\,\epsilon'_{\txant+1}\,\cdots\,\epsilon'_{\cohtime}]}\}
\end{align*}
where $\epsilon'_1,\dots, \epsilon'_{l}$ and $\epsilon'_{\txant+1},\dots, \epsilon'_{\cohtime}$ are positive real numbers chosen so that the diagonal elements of $\matLambda'_{\epsilon}$ are distinct.
Let also $\altmatSigma_{\epsilon}\define\diag\{\tp{[\sigma^2_1\,\cdots\,\sigma^2_{\rxant}\,\,\epsilon_{\rxant+1}\,\cdots\,\epsilon_{\cohtime}]}\}$, where $\epsilon_{\rxant+1},\dots,\epsilon_{\cohtime}$ are positive real numbers chosen so that the diagonal elements of $\altmatSigma_{\epsilon}$ are distinct.
To obtain $\genericpdf_{\matY\given\matD}$, we need to replace~\eqref{eq:reduction_to_itzykson_zuber} with (\ref{eq:reduction_to_itzykson_zuber_v2}) on the top of next page,\addtocounter{equation}{1}
%
%
\newcounter{tempequationcounter}
\begin{figure*}[!t]
\normalsize
\setcounter{tempequationcounter}{\value{equation}}
\begin{IEEEeqnarray}{rCl}
\setcounter{equation}{59}
\label{eq:reduction_to_itzykson_zuber_v2}
	\int_{\unitgroup(\cohtime)}e^{\tr\lefto(\altmatSigma\matPhi\matLambda\herm{\matPhi}\right)} d\altmatPhi
	=\lim_{\epsilon_{\rxant+1}\to 0,\dots, \epsilon_{\cohtime}\to 0} \,\,\,
	\lim_{\epsilon'_{1}\to 0,\dots,\epsilon'_{l}\to 0} \,\,\,
	\lim_{\epsilon'_{\txant+1}\to 0,\dots,\epsilon'_{\cohtime}\to 0}   \int_{\unitgroup(\cohtime)}e^{\tr\lefto(\altmatSigma_{\epsilon}\altmatPhi\matLambda'_{\epsilon}\herm{\altmatPhi}\right)} d\altmatPhi
\end{IEEEeqnarray}
\setcounter{equation}{\value{tempequationcounter}}
\hrulefill
\end{figure*}
%
and then follow the same steps leading to~\eqref{eq:conditional-pdf-Y-closed-form}.
The corresponding steps are omitted.
For simplicity, in the remainder of the proof we shall focus exclusively on the case $\cohtime\leq\rxant$.


\subsection{Step 2} 
\label{sec:step_2}

\subsubsection{Computing~\pdfaltsigmacond} 
\label{par:computing}


To obtain $\pdfaltsigmacond$ from $\genericpdf_{\matY\given \matD}$, we express \matY in terms of its SVD [see~\eqref{eq:SVD-Y}], which yields
\begin{IEEEeqnarray}{rCl}
   \IEEEeqnarraymulticol{3}{l}{\genericpdf_{\matU,\matSigma,\matV \given \matD}(\matU,\matSigma,\matV\given \matD)}\notag\\
   \quad&=&\genericpdf_{\matY\given\matD}(\matU\matSigma\herm{\matV}\given\matD)\cdot J_{\rxant,\cohtime}(\sigma_1,\cdots,\sigma_{\cohtime})\label{eq:conditional_pdf_in_SVD_coordinates}
\end{IEEEeqnarray}
where $J_{\rxant, \cohtime}$ is the Jacobian of the SVD transformation given in~\eqref{eq:jacobian} (recall that we assumed $\cohtime \leq \rxant$, and, hence, $\minbound = \cohtime$ and $\maxbound = \rxant$  ).

Next, we integrate the RHS of~\eqref{eq:conditional_pdf_in_SVD_coordinates} over $\matU$ and $\matV$ and then operate the change of variable $\vecsigma \mapsto \altvecsigma$ defined in~\eqref{eq:normalized_singular_values}.
These two steps yield
\begin{IEEEeqnarray}{rCl}
\IEEEeqnarraymulticol{3}{l}{   	\pdfaltsigmacond(\altvecsigma \given \matD)}\notag\\
\,\,&=& \frac{2^\cohtime e^{ - \sum\nolimits_{i=\txant+1}^{\cohtime}\altsigma_i^2}}{\prod\limits_{i=\rxant-\cohtime+1}^{\rxant} \Gamma(i) \cdot   \prod\limits_{i=1}^{\cohtime-\txant} \Gamma(i)}
\cdot\frac{\det\lefto(\altsnr^{-1}\matD^{-2} + \matI_\txant\right)^{\cohtime-\txant} }{\det\lefto(\matD^2 + \altsnr^{-1}\matI_\txant\right)^{\rxant-\txant+1}}
\notag\\
&&\cdot \prod\limits_{i<j}^{\txant}\bigl(\altsigma_i^2 - \altsigma_j^2\bigr) \cdot\prod\limits_{\txant<i<j}^{\cohtime}\!\!\!(\altsigma_i^2 - \altsigma_j^2)
\cdot\prod\limits_{i=1}^{\txant} \prod\limits_{j=\txant+1}^{\cohtime}
\left(\altsigma_i^2 - \frac{\altsigma_j^2}{\altsnr}\right)
\notag\\
&&\cdot\,\underbrace{e^{-\altsnr\sum\nolimits_{i=1}^{\txant}\altsigma_i^2}\det(\matM)}_{\define c_\snr(\altvecsigma)}{}
\cdot\,\frac{\prod\limits_{i=1}^{\cohtime}\altsigma_i^{2(\rxant-\cohtime)+1}}{\prod\limits_{i<j}^{\txant}(d_i^2-d_j^2)}
.\label{eq:cond_normalized_singular_value_pdf}
\end{IEEEeqnarray}
%
%
%
%

\subsubsection{Convergence of \pdfaltsigmacond to \pdfucond as $\snr \to \infty$}
We start by characterizing the limit ${\snr\to\infty}$ of $c_{\snr}(\altvecsigma)$ in~\eqref{eq:cond_normalized_singular_value_pdf}.
Let \matL be a $\cohtime \times \cohtime$ matrix defined as in (\ref{eq:matrix-L-definition}) on the top of next page.
\begin{figure*}[!t]
\normalsize
\setcounter{tempequationcounter}{\value{equation}}
\begin{IEEEeqnarray}{rCl}
\setcounter{equation}{62}
\matL &\define& \matM\cdot\diag\{\tp{[e^{-\altsnr\tilde{\sigma}_1^2}\,\cdots\,e^{-\altsnr\tilde{\sigma}_\txant^2}\,\underbrace{1\, \ldots\,1}_{\cohtime-\txant}]}\}\notag\\
&=& \left(
          \begin{array}{ccc|ccc}
            e^{-\frac{\altsnr}{1+\altsnr d_1^2}\tilde{\sigma}_1^2 } & \cdots & e^{-\frac{\altsnr}{1+\altsnr d_1^2}\tilde{\sigma}_\txant^2 } & e^{\lambda_1\tilde{\sigma}_{\txant+1}^2} & \cdots & e^{\lambda_1\tilde{\sigma}_{\cohtime}^2} \\
            \vdots &\ddots  & \vdots & \vdots & \ddots & \vdots \\
            e^{-\frac{\altsnr}{1+\altsnr d_\txant^2}\tilde{\sigma}_1^2 } & \cdots & e^{-\frac{\altsnr}{1+\altsnr d_\txant^2}\tilde{\sigma}_\txant^2 } & e^{\lambda_\txant\tilde{\sigma}_{\txant+1}^2 } & \cdots & e^{\lambda_\txant\tilde{\sigma}_\cohtime^2 }\\
            \hline
            (\altsnr\altsigma_1^2)^{\cohtime
            -\txant-1}\cdot e^{-\altsnr \tilde{\sigma}_1^2} & \cdots & (\altsnr\altsigma_\txant^2)^{\cohtime-\txant-1}\cdot e^{-\altsnr \tilde{\sigma}_\txant^2} & \tilde{\sigma}_{\txant+1}^{2(\cohtime-\txant-1)} & \cdots & \tilde{\sigma}_\cohtime^{2(\cohtime
            -\txant-1)} \\
            \vdots & \ddots & \vdots & \vdots & \ddots & \vdots \\
            e^{-\altsnr \tilde{\sigma}_1^2} & \cdots &  e^{-\altsnr \tilde{\sigma}_\txant^2} & 1 & \cdots & 1  \\
          \end{array}
        \right)\notag\\
        &=& \left(\begin{array}{cc}
        \matL_{11} &\matL_{12}\\
        \matL_{21} &\matL_{22}\\
        \end{array}\right).\label{eq:matrix-L-definition}
\end{IEEEeqnarray}
\setcounter{equation}{\value{tempequationcounter}}
\hrulefill
\end{figure*}
\addtocounter{equation}{1}
Observe now that $c_{\snr}(\altvecsigma)=\det(\matL)$ and that $\matL_{21}$ vanishes as $\snr \to \infty$.
These two facts imply that
\begin{IEEEeqnarray}{rCl}
\lim\limits_{\snr\rightarrow\infty} c_\snr(\altvecsigma) &=&
\lim\limits_{\snr\rightarrow\infty} \det(\matL)\notag\\
 &=& \lim\limits_{\snr\rightarrow\infty}\Bigl(\det(\matL_{11}) \det(\matL_{22})\Bigr)\notag\\
& =&\det\bigl(\widetilde{\matL}\bigr)\cdot \prod\limits_{\txant<i<j}^{\cohtime}(\tilde{\sigma}_i^2-\tilde{\sigma}_j^2)\label{eq:limit-c-sigma}
\end{IEEEeqnarray}
with $\widetilde{\matL}$ being a $\txant \times \txant$ matrix defined by $\bigl[\widetilde{\matL}\bigr]_{i,j}=e^{-\altsigma_j^2/d_i^2}$.
Substituting (\ref{eq:limit-c-sigma}) into (\ref{eq:cond_normalized_singular_value_pdf}), we get after some algebraic manipulations
%
%
%
%
%
%
\begin{align}
&\lim\limits_{\snr\rightarrow\infty}\pdfaltsigmacond(\altvecsigma \given \matD) \notag\\
&\,=
\underbrace{
\frac{2^\txant\det\bigl(\widetilde{\matL}\bigr)\cdot\prod\limits_{i=1}^{\txant} \tilde{\sigma}_i^{2(\rxant-\txant)+1}}{\det\lefto(\matD\right)^{2\rxant}  \cdot\!\! \prod\limits_{i=\rxant-\txant+1}^{\rxant}\!\! \Gamma(i)}\cdot \biprod{i<j}{\txant}\frac{\tilde{\sigma}_i^2-\tilde{\sigma}_j^2}{d_i^2-d_j^2}d_i^2 d_j^2
}_{\define\genericpdf_1(\altsigma_1,\dots,\altsigma_\txant)}{}\notag\\
&\,\quad\cdot \underbrace{2^{\cohtime-\txant} e^{-\sum\nolimits_{i=\txant+1}^{\cohtime}\!\!\tilde{\sigma}_i^2}  \cdot\frac{\prod \limits_{i=\txant+1}^\cohtime \!\!\!\tilde{\sigma}_i^{2(\rxant-\cohtime)+1}\cdot\!\!\!\prod\limits_{\txant < i<j}^{\cohtime} \!\!\!\left( \tilde{\sigma}_i^2-\tilde{\sigma}_j^2\right)^2}{\prod\limits_{i=1}^{\cohtime - \txant}\Gamma(i) \cdot\prod\limits_{i=\rxant-\cohtime + 1}^{\rxant - \txant}\Gamma(i)}}_{\define\genericpdf_2(\altsigma_{\txant+1},\dots,\altsigma_{\cohtime})}{}\label{eq:asymptotics_conditional_pdf}
\end{align}
The proof of part~2 is concluded by noting that
\begin{IEEEeqnarray*}{rCl}
\lim\limits_{\snr\rightarrow\infty}\pdfaltsigmacond(\altvecsigma \given \matD) &=&\genericpdf_1(\altsigma_1,\dots,\altsigma_\txant) \cdot \genericpdf_2(\altsigma_{\txant+1},\dots,\altsigma_{\cohtime})\\
&=& 	\pdfucond(\altvecsigma\given\matD)
\end{IEEEeqnarray*}
where the last equality follows from~\cite[Thms.~2.17 and~2.18]{tulino04a}.
\subsection{Step 3} 
\label{sec:part_3}
We next establish that the function $\pdfaltsigmacond(\altvecsigma\given\matD)\cdot  \pdfoptDpar(\matD)$ is bounded. By \fref{eq:input-norm-distribution}, this is obviously true for the case when $\matD  \in \diagsetK(\snr_0) $. We analyze next the case $\matD\notin \diagsetK(\snr_0)$.
Throughout this appendix, we shall use \constapp to indicate a constant term that does not depend on \altvecsigma, \matD, and \snr.
Note that \constapp can take on different values at each appearance.
We start by observing that for an arbitrary $\zeta \in (0,1)$, there exists a
$\snrth>0$ such that  $\Prob\{\matD \in \widetilde{\setK}(\snr_0) ,\matD\distas\distoptD\}\leq \zeta$   for all $\snr>\snrth$.
Hence, for $\snr>\snrth$,
\begin{IEEEeqnarray}{rCl}
\label{eq:q_opt_snr}
\pdfoptDpar(\matD) = \frac{\pdfoptD(\matD)}{1-\Prob\{\matD \in \widetilde{\setK}(\snr_0) , \matD\distas\distoptD\}}
\leq  \frac{\pdfoptD(\matD)}{1-\zeta}.\IEEEeqnarraynumspace
\end{IEEEeqnarray}
Since we are interested in the limit $\snr\to\infty$, we will assume throughout that $\snr>\snrth$,  so that~\eqref{eq:q_opt_snr} holds.
Let $d^2\define\cohtime\rxant/\minparam$.
It follows from~\eqref{eq:pdf_beta_nonsingular} and from the change of variable theorem that
\begin{IEEEeqnarray}{rCl}
\pdfoptD(\matD)&=&\genericpdf_{\lambda_1,\ldots,\lambda_\txant}\lefto(\frac{d_1^2}{d^2},\cdots,\frac{d_{\txant}^2}{d^2}\right)\cdot \prod_{i=1}^{\txant}\frac{2d_i}{d}  \notag\\
	&=&\constapp \cdot  \prod\limits_{i=1}^{\txant}\left(\frac{d_i}{d}\right)^{2(\cohtime-2\txant)+1}\left(1-\frac{d_i^2}{d^2}\right)^{\rxant-\cohtime}\notag\\
	&& \cdot \biprod{i<j}{\txant}\left(\frac{d_i^2-d_j^2}{d^2}\right)^2.\label{eq:a-priori-prob_expression}
\end{IEEEeqnarray}
Here, the second equality follows by setting  $m=\txant$, $p=\cohtime-\txant$, and $n=\txant+\rxant-\cohtime$ in~\eqref{eq:pdf_beta_nonsingular}.
Substituting~\eqref{eq:a-priori-prob_expression} into~\eqref{eq:q_opt_snr} we obtain
\begin{IEEEeqnarray}{rCl}
	 \IEEEeqnarraymulticol{3}{l}{\pdfaltsigmacond(\altvecsigma\given\matD)\cdot  \pdfoptDpar(\matD)}\notag\\
  \, &\leq& \constapp \cdot
	\frac{ \det \lefto(\altsnr^{-1}\matD^{-2} + \matI_\txant\right)^{\cohtime-\txant} }{\det\lefto(\matD^2 + \altsnr^{-1}\matI_\txant\right)^{\rxant-\txant+1}} 
\cdot\det(\matM) e^{-\altsnr\sum\nolimits_{i=1}^{\txant}\tilde{\sigma}_i^2}
\notag\\
	&& \cdot\, \underbrace{\prod\limits_{i=1}^{\txant}\tilde{\sigma}_i^{2(\rxant-\cohtime)+1}\cdot \prod\limits_{i<j}^{\txant}(\tilde{\sigma}_i^2 - \tilde{\sigma}_j^2)\cdot
	\prod\limits_{i=1}^{\txant} \prod\limits_{j=\txant+1}^{\cohtime}\!\!\left(\tilde{\sigma}_i^2 - \frac{\tilde{\sigma}_j^2}{\altsnr}\right)}_{\leq \prod\limits_{i=1}^{\txant} \tilde{\sigma}_i^{2(\rxant-i)+1}}{}\notag\\
&&  \cdot\, e^{-\!\!\sum\nolimits_{i=\txant+1}^{\cohtime}\!\tilde{\sigma}_i^2}
\cdot\underbrace{\prod\limits_{i=\txant+1}^{\cohtime}\!\!\tilde{\sigma}_i^{2(\rxant-\cohtime)+1}\cdot \!\!\prod\limits_{\txant<i<j}^{\cohtime}\!\!\!(\tilde{\sigma_i}^2 - \tilde{\sigma_j}^2)}_{\leq \prod\limits_{i=\txant+1}^{\cohtime} \tilde{\sigma}_i^{2(\rxant-i)+1}}{}\notag\\
	&& \cdot\, \underbrace{\prod\limits_{i<j}^{\txant}(d_i^2-d_j^2) \cdot \prod\limits_{i=1}^{\txant}d_i^{2(\cohtime-2\txant)+1} (d^2-d_i^2)^{\rxant-\cohtime}}_{\leq \constapp \cdot \prod \limits_{i=1}^{\txant} d_i}{}     \notag\\
	%
	&\leq& \constapp \cdot \underbrace{\prod\limits_{i=1}^{\txant}\Bigl[1+1/(\altsnr d_i^2)\Bigr]^{\cohtime -\txant}}_{\leq(1+\txant/\snr_0)^{\txant(\cohtime-\txant)}=\constapp}{}  \cdot \frac{\det(\matM)\prod\limits_{i=1}^{\cohtime} \tilde{\sigma}_i^{2(\rxant-i)+1} }{\prod\limits_{i=1}^{\txant} (\altsnr^{-1}+d_i^2)^{\rxant-\txant+1/2}}\notag \\
&& \cdot\underbrace{ \prod \limits_{i=1}^{\txant} \frac{d_i}{(\altsnr^{-1}+d_i^2)^{1/2}}}_{\leq 1}{}
	 \cdot\exp\lefto(-\altsnr\sum\limits_{i=1}^{\txant}\tilde{\sigma}_i^2- \!\!\sum\limits_{i=\txant+1}^{\cohtime}\tilde{\sigma}_i^2\right)\notag\\
	&\leq& \constapp \cdot \det\lefto(\matN\right)\label{eq:app-bounding-f-q}
\end{IEEEeqnarray}
where the $\cohtime \times \cohtime$ matrix \matN is defined as follows:
\begin{IEEEeqnarray*}{rCl}
\matN &\define& \diag\biggl\{\Bigl[(\snrnorm^{-1} + d_1^2)^{-(\rxant-\txant+1/2)}\, \notag\\
&&\hfill\cdots\, (\snrnorm^{-1} + d_\txant^2)^{-(\rxant-\txant+1/2)},
\underbrace{1\,\,\cdots\,\,1}_{\cohtime-\txant}{}\Bigr]^{\mathsf{T}}\biggr\}\notag\\
&& \cdot\, \matM  \cdot \diag\lefto\{\tp{\lefto[e^{-\snrnorm \tilde{\sigma}_1^2}\,\, \cdots\,\,e^{-\snrnorm \tilde{\sigma}_\txant^2}\,\, e^{- \tilde{\sigma}_{\txant+1}^2}\,\, \cdots\,\, e^{- \tilde{\sigma}_{\cohtime}^2}\right]}\right\}\\
&& \cdot \diag\lefto\{\tp{\lefto[ \tilde{\sigma}_1^{2(\rxant-1)+1}\,\,\cdots\,\, \tilde{\sigma}_\cohtime^{2(\rxant-\cohtime)+1}\right]}\right\}.
\end{IEEEeqnarray*}
Next, we upper-bound $\det(\matN)$ by bounding the entries $n_{i,j}$ of \matN, which are given by
%
%
%
%
\begin{equation*}
n_{i,j} = \left\{
\begin{split}
&\frac{\exp\lefto[-\left(d_i^2+\snrnorm^{-1}\right)^{-1}\tilde{\sigma}_j^2\right]}{\left(d_i^2+\snrnorm^{-1}\right)^{\rxant-\txant+1/2}}
\cdot \tilde{\sigma}_j^{2(\rxant-j)+1},\\
&\qquad\qquad\qquad\quad \qquad \quad\,\, 1\leq i\leq \txant,\,\,    1\leq j \leq \txant\\[2mm]
&\frac{\exp\lefto[-(1+\snrnorm d_i^2)^{-1}\tilde{\sigma}_j^2\right]} {\left(d_i^2+\snrnorm^{-1}\right)^{\rxant-\txant+1/2}}
\cdot \tilde{\sigma}_j^{2(\rxant-j)+1},\\
&\qquad\qquad\qquad\quad \qquad\quad\,  1\leq i\leq \txant,\,\, \txant < j\leq \cohtime\\[2mm]
&\left(\snrnorm \tilde{\sigma}_j^2\right)^{\cohtime - i}\cdot e^{-\snrnorm \tilde{\sigma}_j^2}\cdot \tilde{\sigma}_j^{2(\rxant - j)+1},\\
&\qquad\qquad\qquad\quad \qquad\quad\,   \txant  <i\leq \cohtime,\,\, 1\leq j \leq \txant\\[2mm]
&\tilde{\sigma}_j^{2(\cohtime +\rxant - i-j)+1} \cdot e^{-\tilde{\sigma}_j^2},\,\,  \txant<i\leq\cohtime,\,\,
  \txant<j\leq\cohtime.
\end{split}
\right.
\end{equation*}

%
In the following, we shall repeatedly make use of the fact that the function $f(x)=e^{-\beta x^2}x^{\alpha}$ with $\alpha,\beta,x>0$  is maximized for $x=x^*=\sqrt{\alpha/(2\beta)}$, and that the corresponding maximum values is $f(x^*)=\bigl[\alpha/(2\beta e)\bigr]^{\alpha/2}$.
This implies that
\begin{align}\label{eq:useful-inequality}
f(x)=e^{-\beta x^2} x^{\alpha} \leq \left(\frac{\alpha}{2 \beta e}\right)^{\alpha/2}, \quad x>0.
\end{align}
\subsubsection{Case $1\leq i\leq \txant,\,\,\,  1\leq j\leq \txant$} 
\label{sec:case_1}
\begin{IEEEeqnarray*}{rCl}
    n_{i,j} &\stackrel{(a)}{\leq}& \constapp \cdot \left(d_i^2+\snrnorm^{-1}\right)^{\rxant-j+1/2} \left(d_i^2+\snrnorm^{-1}\right)^{-(\rxant-\txant+1/2)} \\
    &=&\constapp \cdot\left(d_i^2+\snrnorm^{-1}\right)^{\txant-j}\\
	&\stackrel{(b)}{\leq}&\constapp \cdot \left(d^2+\frac{\txant}{\snrth}\right)^{\txant-j} =\constapp.
\end{IEEEeqnarray*}
Here, (a) follows from (\ref{eq:useful-inequality}) by setting $\alpha = 2(\rxant -j)+1$, $\beta = (d_i^2+\tilde{\snr}^{-1} )^{-1} $, and $x= \tilde{\sigma}_j$;~(b) follows because $d_i\leq d$ and $j\leq \txant$.
\subsubsection{Case $1\leq i\leq \txant,\,\,\, \txant< j\leq \cohtime$} 
\label{sec:case_2}
\begin{IEEEeqnarray}{rCl}
 n_{i,j}&=&\underbrace{ \exp\lefto[-(1+\snrnorm d_i^2)^{-1}\tilde{\sigma}_j^2\right]}_{\leq 1}{} \cdot \tilde{\sigma}_j^{2(\rxant-j)+1}\notag\\
 && \cdot \underbrace{\left(d_i^2+\snrnorm^{-1}\right)^{-(\rxant-\txant+1/2)}}_{\stackrel{(a)}{\leq} \left(\frac{\snrnorm}{ 1+\snr_0/\txant} \right)^{\rxant-\txant+1/2} }{}\notag\\
  &\leq& \constapp \cdot \tilde{\sigma}_j^{2(\rxant-j)+1}  \snrnorm^{\rxant-\txant+1/2}.\label{eq:app-bounding-N-case-2}
 \end{IEEEeqnarray}
Here,~(a) follows because $d_i^2 \snr \geq \snr_0$.
\subsubsection{Case $\txant<i\leq \cohtime,\,\,\, 1\leq j\leq \txant$} 
\label{sec:case_3}
\begin{IEEEeqnarray*}{rCl}
 n_{i,j} &=& \snrnorm^{\cohtime -i } e^{-\snrnorm \tilde{\sigma}_j^2} \tilde{\sigma}_j^{2(\cohtime+\rxant-i-j)+1}\\
 &\stackrel{(a)}{\leq}& \constapp \cdot \snrnorm^{\cohtime -i }\cdot \left(\frac{1}{\snrnorm}\right)^{\cohtime+\rxant-i-j+1/2}\\
 &=& \constapp \cdot \snrnorm^{-(\rxant-j+1/2)}\\
 &\leq& \constapp \cdot \snrth^{-(\rxant-j+1/2)}=\constapp.
\end{IEEEeqnarray*}
Here, (a) follows from (\ref{eq:useful-inequality}) by setting $\alpha = 2(\cohtime + \rxant -i -j)+1$, $\beta = \snrnorm $, and $x= \tilde{\sigma}_j$.
\subsubsection{Case $\txant<i\leq\cohtime,\,\,\, \txant<j\leq\cohtime$} 
\label{sec:case}
We have $n_{i,j}\leq \constapp$, which follows directly from~(\ref{eq:useful-inequality}) by setting $\alpha = 2(\cohtime+\rxant-i-j)+1$, $\beta =1$, and $x=\tilde{\sigma_j}$.





To show that $\det(\matN)$ is bounded, it remains to further analyze case 2, where  $n_{i,j}$ is not bounded.
Let $\{i_1,\ldots,i_\cohtime\}$ be  an arbitrary permutation of $\{1,\ldots, \cohtime\}$.
Then~\cite[Sec.~0.3]{horn85a}
\begin{align}\label{eq:det_expression_sign}
	\det(\matN) = \sum\limits_{(i_1,\ldots,i_\cohtime)} \mathrm{sgn}(i_1,\ldots,i_\cohtime) \cdot n_{i_1,1}\cdot \ldots \cdot n_{i_\cohtime,\cohtime}.
\end{align}
Here, the sum is over all the $(\cohtime!)$ permutations of $\{1,\ldots, \cohtime\}$ and $\mathrm{sgn}(\cdot)$ denotes the sign of the permutation~\cite[p.~8]{horn85a}.
We observe that for each $n_{i,j}$ ($1\leq i\leq \txant, \txant <j\leq \cohtime$) that appears in the product on the RHS of~\eqref{eq:det_expression_sign}, there exists a factor $n_{i',j'}$ in the same product with $\txant<i'\leq \cohtime$, $1\leq j'\leq \txant$ (case 3).
Note now that
\begin{IEEEeqnarray*}{rCl}
\IEEEeqnarraymulticol{3}{l}{n_{i,j} \cdot n_{i',j'} }\notag\\
\,\, &\stackrel{(a)}{\leq}&  \constapp \cdot\underbrace{\tilde{\sigma}_j^{2(\rxant-j)+1}}_{\stackrel{(b)}{\leq}\left(\snrnorm \tilde{\sigma}_{j'}^2\right)^{\rxant-j+1/2} }{} \cdot \snrnorm^{\rxant-\txant+1/2}\notag\\
&&\cdot\,  \left(\snrnorm \tilde{\sigma}_{j'}^2\right)^{\cohtime - {i'}} \cdot e^{-\snrnorm \tilde{\sigma}_{j'}^2}\cdot \tilde{\sigma}_{j'}^{2(\rxant - j')+1} \\
&\leq& \constapp \cdot \snrnorm^{2\rxant+\cohtime-\txant+1-j-i'}\cdot \underbrace{\tilde{\sigma}_{j'}^{2(2\rxant+\cohtime+1-j-i'-j')} e^{-\snrnorm\tilde{\sigma}_{j'}^2}}_{\stackrel{(c)}{\leq} \constapp \cdot \tilde{\snr}^{-(2\rxant+\cohtime +1 -j -i'-j')}}{}\\
&\leq& \constapp \cdot \tilde{\snr}^{-(\txant-j')} \\
&\leq& \constapp \cdot \snrth^{-(\txant-j')}=\constapp.
\end{IEEEeqnarray*}
Here, (a) follows from \fref{eq:app-bounding-N-case-2}, in (b) we used that $\snrnorm \tilde{\sigma}_{j'}^2=\sigma_{j'}^2\geq \sigma_j^2 =\tilde{\sigma}_j^2$ for $j'\leq \txant<j$, and (c) follows from (\ref{eq:useful-inequality}) by setting $\alpha =2(2\rxant+\cohtime+1-j-i'-j') $, $\beta=\snrnorm$, and $x=\tilde{\sigma}_{j'}$.

Summarizing, we showed that
\begin{align*}
\abs{\pdfaltsigmacond(\altvecsigma\given\matD)\cdot  \pdfoptDpar(\matD)} \leq \constapp\cdot \det(\matN)\leq \constapp
\end{align*}
which concludes the proof of the lemma.




\section{Proof of Lemma~\ref{lem:useful_asyptotic_results}} 
\label{app:proof_of_lemma_lem:useful_asyptotic_results}
Throughout this appendix, we shall set $d\define\sqrt{\rxant\cohtime/\minparam}$,  and   $\snrnorm\define\snr/\txant$, and denote by $\pdfoptDpar$ and $\pdfoptD$ the pdfs corresponding to the probability distributions \distoptDpar and \distoptD, respectively,
by \pdfaltsigma and \pdfu the pdf of \altvecsigma and \vecu, respectively, and by \pdfaltsigmacond and \pdfucond  the conditional pdf of \altvecsigma and \vecu given \matD, respectively.
We shall use \constapp to denote a finite constant; its value might change at every appearance.
Since the lemma only addresses limiting behaviors as $\snr\to \infty$, we shall assume throughout that $\snr>\snrth>0$.
Finally, for simplicity  we shall focus exclusively  on the case $\cohtime\leq \rxant$; the proof for the case $\cohtime>\rxant$ follows from analogous steps.
%
%

\subsection{Proof of Part 1} 
\label{sec:part_1}
   The proof is based on the following theorem.
\begin{thm}[\!\protect{\cite[Thm.~1]{godavarti04-01a}}]\label{thm:convergence_in_differential_entropy}
   Let $\{\vecx_i \in \complexset^m \}$ be a sequence of random vectors with pdfs ${\mathsf{f}_i}$ and let $\vecx \in \complexset^{m}$ be a random vector with pdf $\mathsf{f}$.
Assume that $\mathsf{f}_i$ converges to $ \mathsf{f}$ pointwise.
If there exist
\begin{inparaenum}[i)]
	\item a finite constant $F>0$ such that $\max\{ \sup_\vecx \mathsf{f}_i(\vecx), \sup_\vecx\mathsf{f}(\vecx)\}\leq F$ for all $i$, and
	\item a finite constant $L>0$ such that $\max\{\int\|\vecx\|^\kappa \mathsf{f}_i(\vecx) d\vecx,\int\|\vecx\|^\kappa \mathsf{f}(\vecx) d\vecx \}\leq L$ for some $\kappa >1$ and all $i$,
\end{inparaenum}
then $\difent(\vecx_i)\rightarrow\difent(\vecx)$.
\end{thm}

Since we established in~\fref{app:proof_of_lemma_lem:convergence_in_pdf} that $\pdfaltsigma$ converges to \pdfu pointwise as $\snr \to \infty$, we just need to verify that both \pdfaltsigma and \pdfu satisfy the conditions i) and ii) in~\fref{thm:convergence_in_differential_entropy}.
\subsubsection{\pdfaltsigma  and \pdfu are bounded} 
\label{sec:boundness}
Because of~\eqref{eq:bounded_by_const_app1}, and since $0<d_i\leq d$,  we have that
\begin{align*}
	 \pdfaltsigma(\altvecsigma)=\int \pdfaltsigmacond(\altvecsigma | \matD) \pdfoptDpar(\matD)d\matD \leq \constapp \cdot d^{\txant}=\constapp.
\end{align*}
To show that \pdfu is bounded, we first prove that $\pdfucond\cdot \pdfoptD$ is bounded by using~\eqref{eq:asymptotics_conditional_pdf} and~\eqref{eq:a-priori-prob_expression}:
\begin{IEEEeqnarray*}{rCl}
\IEEEeqnarraymulticol{3}{l}{\pdfucond(\vecu\given\matD)\cdot \pdfoptD(\matD)}\\
 \,\,&=&
 \constapp \cdot \frac{\det(\widetilde{\matL})}{\prod\limits^{\txant}_{i=1}d_i^{2(\rxant-\txant+1)}}\cdot \underbrace{\prod\limits_{i=1}^{\txant
    } u_i^{2(\rxant-\txant)+1}\cdot\prod\limits_{i<j}^{\txant}\left(u_i^2-u_j^2\right)}_{\leq \prod\limits_{i=1}^{\txant}u_i^{2(\rxant -i)+1}}{}\\
    && \cdot\, e^{-\sum\nolimits_{i=\txant+1}^\cohtime u_i^2} \!\cdot \underbrace{\prod \limits_{i=\txant+1}^\cohtime \!\!\! u_i^{2(\rxant-\cohtime)+1} \cdot \!\! \prod \limits_{\txant <i<j}^{\cohtime}\!\! \left( u_i^2-u_j^2\right)^2}_{\leq \prod \limits_{i=\txant+1}^{\cohtime} u_i^{2(\rxant +\cohtime -2i)+1 }}{}\\
   &&\cdot \underbrace{\prod\limits_{i<j}^{\txant}\left(d_i^2-d_j^2\right) \cdot \left(\prod\limits_{i=1}^{\txant}d_i^{2(\cohtime-2\txant)+1} \left(d^2-d_i^2\right)^{\rxant-\cohtime}\right)}_{\leq \constapp \cdot \prod \limits_{i=1}^{\txant} d_i}\\
&\leq& \constapp \cdot \underbrace{\det(\widetilde{\matL}) \cdot \prod\limits_{i=1}^{\txant} d_i^{-(2(\rxant-\txant)+1)}\cdot \prod
 \limits_{j=1}^{\txant}u_j^{2(\rxant-j)+1}}_{\stackrel{(a)}{\leq}\constapp }{}  \\&&
 \cdot \prod \limits_{i=\txant+1}^{\cohtime}\underbrace{ e^{-u_i^2}  u_i^{2(\rxant +\cohtime -2i)+1 }}_{\stackrel{(b)}{\leq} \constapp}{}\\
    &\leq& \constapp.
\end{IEEEeqnarray*}
Here, (a) follows from~\eqref{eq:useful-inequality} with the choice $\alpha=2(\rxant - j)+1$, $\beta = d_i^{-2}$, and $x=u_j$, as detailed below
\begin{multline*}
e^{-u_j^2/d_i^2} \cdot d_i^{-(2(\rxant-\txant)+1)}\cdot u_j^{2(\rxant-j)+1}\\
\leq \constapp \cdot d_i^{2(\txant-j)} \leq \constapp \cdot d^{2(\txant-j)}=\constapp
\end{multline*}
for all $1\leq i,j\leq \txant$; (b) follows again from~\eqref{eq:useful-inequality}.
Thus
\begin{align*}
	\pdfu(\vecu)=\int \pdfucond(\vecu\given\matD)\cdot \pdfoptD(\matD) d\matD \leq \constapp\cdot d^\txant =\constapp.
\end{align*}
%

\subsubsection{\pdfaltsigma  and \pdfu have finite second moment} 
\label{sec:pdfaltsigma_and_pdfu_have_finite_second_moment}
We take $\kappa=2$, and obtain
\begin{IEEEeqnarray}{rCl}
\label{eq:bound_second_moment}
	 \int \vecnorm{\altvecsigma}^2\pdfaltsigma(\altvecsigma) d\altvecsigma &=& \frac{1}{\altsnr}\sum_{i=1}^{\txant}\Ex{}{\sigma^2_i} + \sum_{i=\txant+1}^{\minbound}\Ex{}{\sigma_i^2} \notag\\
	&\stackrel{(a)}{\leq}& \frac{1}{\altsnr}\rxant\cohtime(\snr+1) + (\rxant-\txant)(\cohtime-\txant)\notag\\
	& \leq & (1+1/\snrth)\txant\rxant\cohtime  + (\rxant-\txant)(\cohtime-\txant) \notag\\
&=& \constapp.
\end{IEEEeqnarray}
Here,~(a) follows from~\eqref{eq:the_second_last_term} and~\eqref{eq:the_last_term}.
Furthermore,
\begin{IEEEeqnarray*}{rCl}
	\int \vecnorm{\vecu}^2\pdfu(\vecu)d\vecu &=&\Ex{}{\tr\lefto(\matD\matH\herm{\matH}\matD\right)} + (\rxant-\txant)(\cohtime-\txant) \\
&=&\constapp.
\end{IEEEeqnarray*}
This concludes the proof.



\subsection{Proof of Part 2} 
\label{sec:part_2}
Let $0<\delta<1$ and let $r$ be a positive integer satisfying $r>1$.
Denote by $\pdfaltsigmascalar$ the pdf of $\altsigma_i$ and  by $\pdfuscalar$ the pdf of $u_i$.
The expectation on \fref{lem:useful_asyptotic_results}--Part-2 can be rewritten as follows:
\begin{IEEEeqnarray}{rCl}
\label{eq:splitting_expectation}
	\lim_{\snr\to\infty}\Ex{\pdfaltsigmascalar}{\log\lefto(x \right)} =
	\lim_{\snr\to\infty}\Bigl\{&&
	\Ex{\pdfaltsigmascalar}{\log\lefto(x\right)\cdot I\{x<\delta\}}  \notag\\
	&& +\,\Ex{\pdfaltsigmascalar}{\log\lefto(x \right)\cdot I\{\delta\leq x\leq r\}}\notag\\
	&&+\,\Ex{\pdfaltsigmascalar}{\log\lefto(x \right)\cdot I\{x>r\}}
	\Bigr\}\IEEEeqnarraynumspace
\end{IEEEeqnarray}
where $I\{\cdot\}$ is the indicator function.
We analyze the three terms on the RHS of~\eqref{eq:splitting_expectation} separately.
For the first term,~\cite[Lemma 6.7(a)]{lapidoth03-10a} and \fref{lem:useful_asyptotic_results}--Part 1 imply that
\begin{align*}
   \lim_{\snr\to\infty}\Ex{\pdfaltsigmascalar}{\log\lefto(x \right)\cdot I\{x<\delta\}}= \epsilon_1(\delta)
\end{align*}
where $\epsilon_1(\delta)\to 0$ as $\delta\to 0$.
For the second term, we have that
\begin{multline*}
	\lim_{\snr\to\infty}\Ex{\pdfaltsigmascalar}{\log\lefto(x \right)\cdot I\{\delta\leq x\leq r\}} \\ =  \Ex{\pdfuscalar}{\log\lefto(x \right)\cdot I\{\delta\leq x\leq r\}}
\end{multline*}
as a consequence of the dominated convergence theorem.
Finally, for the third term we proceed as follows:
\begin{IEEEeqnarray}{rCl}
\label{eq:third_term_expection_logx}
	\Ex{\pdfaltsigmascalar}{\log\lefto(x \right)\cdot I\{x>r\}} &=&
	\sum_{l=r}^{\infty}\int_{l}^{l+1} \pdfaltsigmascalar(x) \log (x) dx \notag\\
	&\stackrel{(a)}{\leq}& \sum_{l=r}^{\infty}\int_{l}^{l+1} \pdfaltsigmascalar(x) \sqrt{x}\, dx  \notag\\
	&\leq&  \sum_{l=r}^{\infty}\sqrt{l+1}\int_{l}^{l+1} \pdfaltsigmascalar(x) \, dx\notag\\
	&\leq&  \sum_{l=r}^{\infty}\sqrt{l+1} \int_{l}^{\infty} \pdfaltsigmascalar(x) \, dx \notag\\
	&\stackrel{(b)}{\leq}&  \sum_{l=r}^{\infty} \left[\sqrt{l+1} \cdot \frac{\Ex{\pdfaltsigmascalar}{x^2}}{l^2}\right]\notag\\
	&\stackrel{(c)}{\leq}&  \sum_{l=r}^{\infty} \left[ \sqrt{l+1} \cdot \frac{ \constapp}{l^2}\right]\notag\\
	&\leq& \sqrt{2}\constapp\sum_{l=r}^{\infty}l^{-3/2}.
\end{IEEEeqnarray}
Here, $(a)$ follows because $\log(x)\leq \sqrt{x}$, $x\geq 1$, (b) follows from Markov's inequality, and (c) is a consequence of~\eqref{eq:bound_second_moment}. Note that \eqref{eq:third_term_expection_logx} holds for all $\snr > \snr_{\mathrm{th}}$. Hence, we have
\begin{IEEEeqnarray*}{rCl}
0\leq \epsilon_2(r)\define \lim\limits_{\snr\to\infty} \Ex{\pdfaltsigmascalar}{\log\lefto(x \right)\cdot I\{x>r\}} \leq \sqrt{2}\constapp\sum_{l=r}^{\infty}l^{-3/2}.
\end{IEEEeqnarray*}
Since $\sum_{l=r}^{\infty}l^{-3/2}$ converges, we can make $\epsilon_2(r)$ arbitrarily close to~$0$ by choosing $r$ sufficiently large.
Summarizing, we showed that
\begin{IEEEeqnarray*}{rCl}
	 \lim_{\snr\to\infty}\Ex{\pdfaltsigmascalar}{\log\lefto(x \right)} &=& \Ex{\pdfuscalar}{\log\lefto(x \right)\cdot I\{\delta\leq x \leq r\}} \notag\\&& +\, \epsilon_1(\delta) + \epsilon_2(r).
\end{IEEEeqnarray*}
The RHS of this equality can be made arbitrarily close to  $\Ex{\pdfuscalar}{\log\lefto(x \right)}$ by choosing $\delta$ sufficiently small and $r$ sufficiently large.
This concludes the proof.

\subsection{Proof of Part 3} 
\label{sec:proof_of_part_3}
To establish the desired result, it is sufficient to show that
\begin{align}
\label{eq:exp-i-minus-j}
\Ex{}{\log(\tilde{\sigma}_i -\tilde{\sigma}_j )}=\Ex{}{\log(u_i -u_j )}+o(1),~\snr\rightarrow \infty
\end{align}
and that
\begin{align}
\label{eq:exp-i-plus-j}
\Ex{}{\log(\tilde{\sigma}_i + \tilde{\sigma}_j )}=\Ex{}{\log(u_i +u_j )}+o(1),~\snr\rightarrow \infty.
\end{align}
\fref{lem:useful_asyptotic_results}--Part~1 implies that, for sufficiently large \snr, $\difent(\altsigma_i-\altsigma_j)>-\infty$ and $\difent(\altsigma_i+\altsigma_j)>-\infty$;
We can now establish~(\ref{eq:exp-i-minus-j}) and (\ref{eq:exp-i-plus-j}) through steps similar to the ones in Part 2.

\subsection{Proof of Part~4} 
\label{sec:proof_of_part_4}
    The proof is analogous to the proof of part~2 and part~3.



\bibliographystyle{IEEEtran}
\bibliography{IEEEabrv,publishers,confs-jrnls,giubib}

\begin{biography}[{\includegraphics[width=1in,height =1.25in,clip,keepaspectratio]{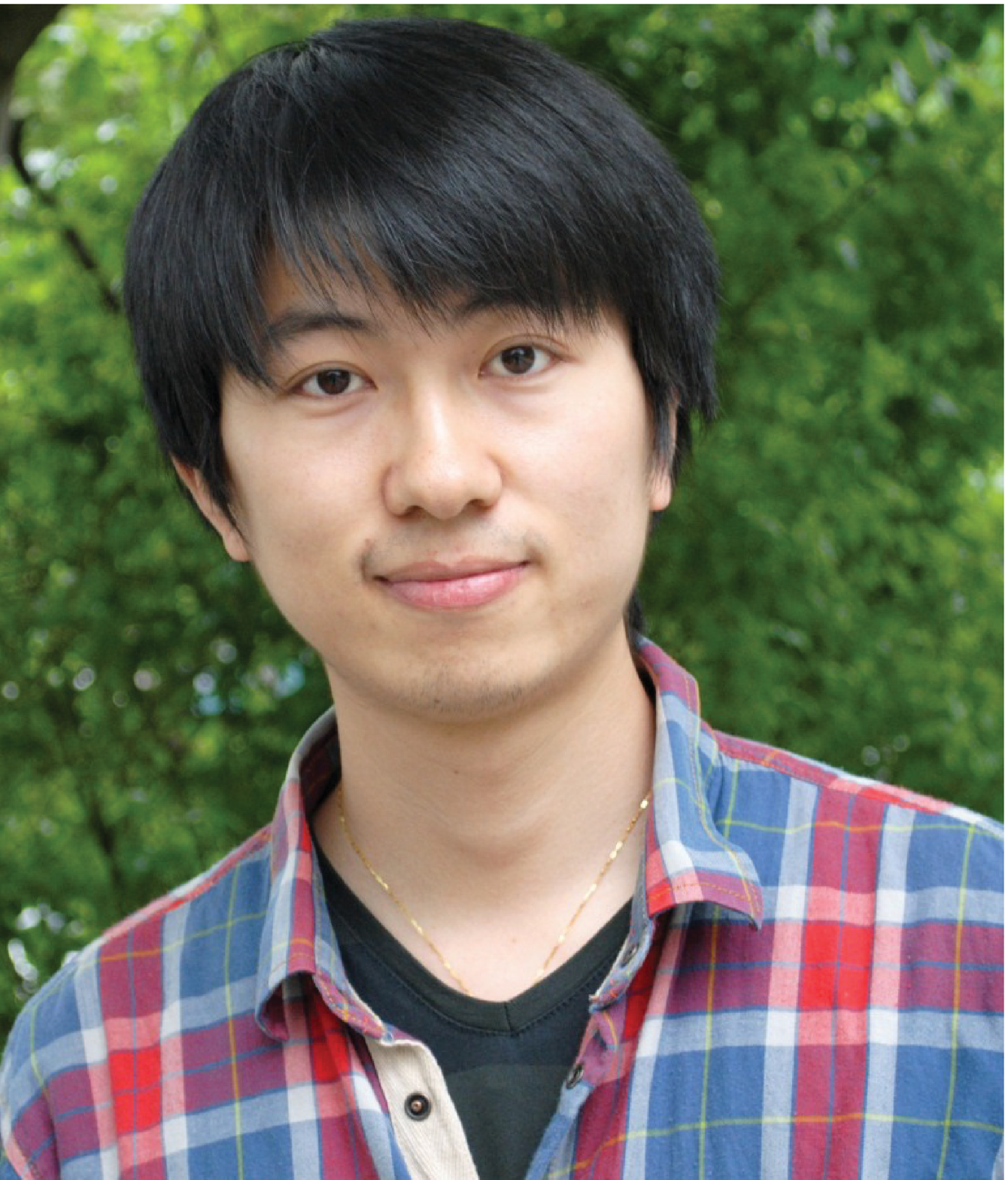}
}]{Wei Yang}(S'09)
received the B.E. degree in communication engineering and M.E. degree in communication and information systems from the Beijing University of Posts and Telecommunications, Beijing, China, in 2008 and 2011, respectively. He is currently pursuing a Ph.D. degree in electrical engineering at Chalmers University of Technology, Gothenburg, Sweden. From July to August 2012, he was a visiting student at the Laboratory for Information and Decision Systems, Massachusetts Institute of Technology, Cambridge, MA.

Mr. Yang is the recipient of a Student Paper Award at the 2012 IEEE International Symposium on Information Theory (ISIT), Cambridge, MA. His research interests are in the areas of information and communication theory.
\end{biography}

\begin{biography}[{\includegraphics[width=1in,height =1.25in,clip,keepaspectratio]{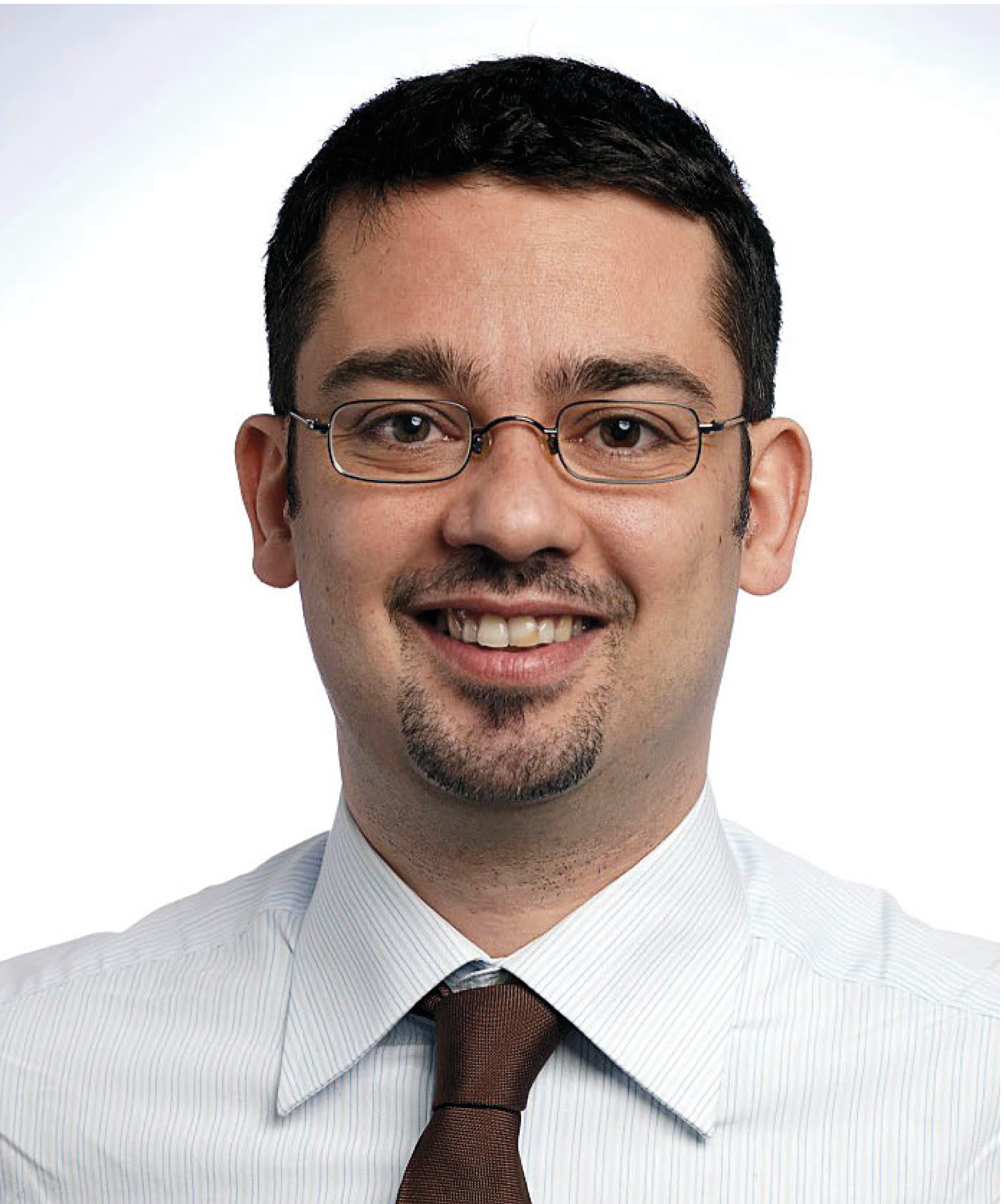}
}]{Giuseppe Durisi}(S'02--M'06--SM'12)
 received the Laurea degree summa cum laude and the Doctor degree both from Politecnico di Torino, Italy, in 2001 and 2006, respectively. From 2002 to 2006, he was with Istituto Superiore Mario Boella, Torino, Italy. From 2006 to 2010 he was a postdoctoral researcher at ETH Zurich, Zurich, Switzerland. Since 2010 he has been an assistant professor at Chalmers University of Technology, Gothenburg, Sweden. He held visiting researcher positions at IMST (Germany), University of Pisa (Italy),  and Vienna University of Technology (Austria).

Dr. Durisi is co-author of a paper that won a student paper award at the International Symposium on Information Theory (ISIT 2012).
He served as TPC member in several IEEE conferences, and is currently publications editor of the IEEE Transactions on Information Theory.
His research interests are in the areas of information theory, communication  theory, and compressive sensing.
\end{biography}

\begin{biography}[{\includegraphics[width=1in,height =1.25in,clip,keepaspectratio]{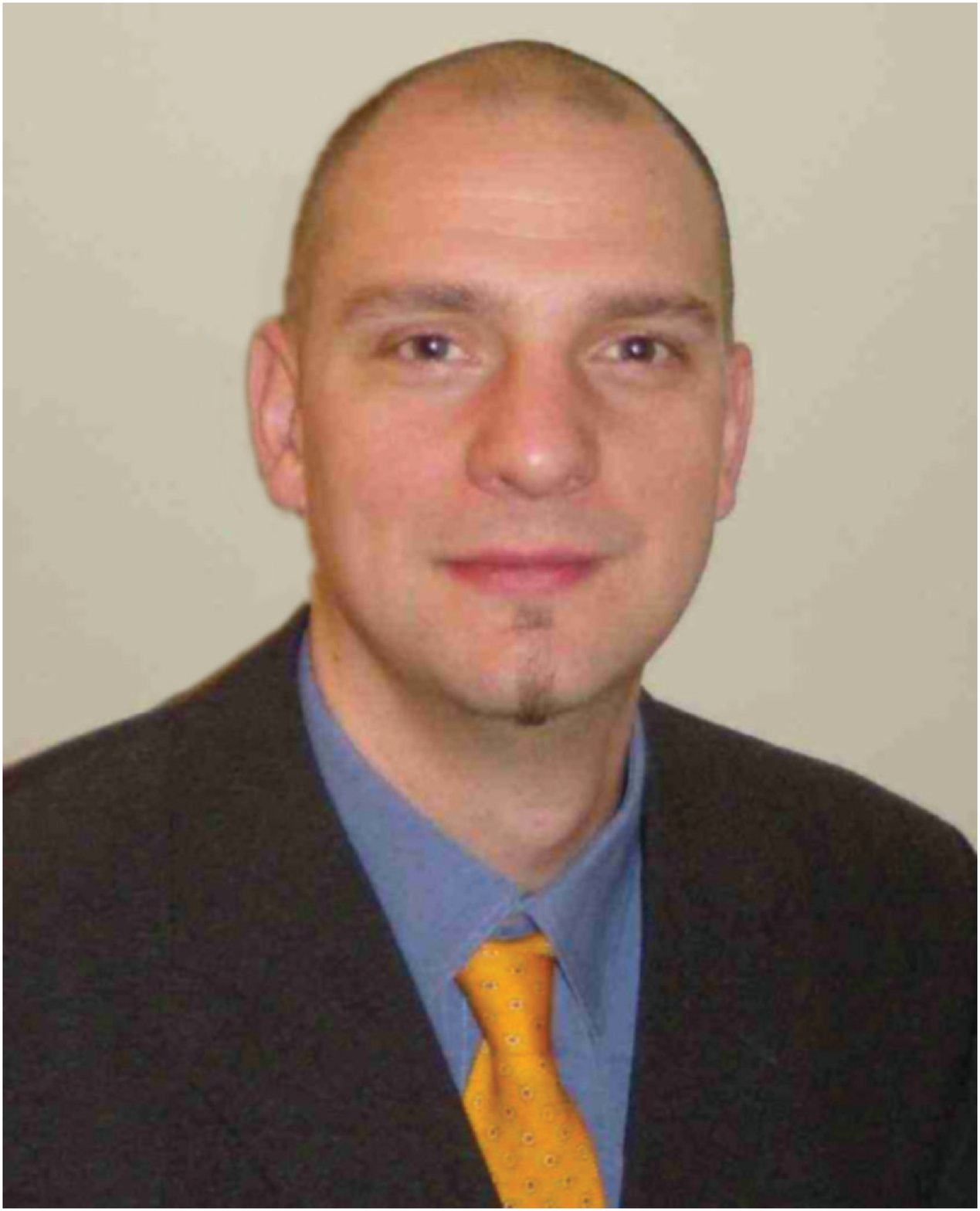}
}]{Erwin Riegler}
(M'07) received the Dipl-Ing. degree in Technical Physics (with distinction) in 2001 and the
Dr. techn. degree in Technical Physics (with distinction) in 2004 from Vienna University of Technology.

He was a visiting researcher at the Max Planck Institute for Mathematics in the Sciences in Leipzig, Germany (Sep. 2004\,--\,Feb. 2005),
the Communication Theory Group at ETH Z\"urich, Switzerland (Sep.~2010\,--\,Feb.~2011 and Jun.~2012\,--\,Nov.~2012),
and the Department of Electrical and Computer Engineering at The Ohio State University in Columbus, Ohio (Mar. 2012).
From 2005 to 2006, he was a post-doctoral fellow at the Institute for Analysis and Scientific Computing, Vienna University of Technology.
From 2007 to 2010, he was a senior researcher at the Telecommunications Research Center Vienna (FTW).
Since 2010, he has been a post-doctoral fellow at the Institute of Telecommunications at Vienna University of Technology.

His research interests include noncoherent communications, machine learning, interference management, large system analysis,
 and transceiver design.
\end{biography}

\end{document}